\documentclass[aps,prd,superscriptaddress,nofootinbib,amsmath,amsfonts,showkeys,showpacs,preprintnumbers,notitlepage,10pt,english]{revtex4-1}
\usepackage{amsmath}
\usepackage{amssymb}
\usepackage{babel}
\usepackage{graphicx}
\usepackage{dcolumn}
\usepackage{bm}
\usepackage[figtopcap]{subfigure}

\makeatletter

\@ifundefined{textcolor}{}
{%
 \definecolor{BLACK}{gray}{0}
 \definecolor{WHITE}{gray}{1}
 \definecolor{RED}{rgb}{1,0,0}
 \definecolor{GREEN}{rgb}{0,1,0}
 \definecolor{BLUE}{rgb}{0,0,1}
 \definecolor{CYAN}{cmyk}{1,0,0,0}
 \definecolor{MAGENTA}{cmyk}{0,1,0,0}
 \definecolor{YELLOW}{cmyk}{0,0,1,0}
 }
\usepackage{enumerate}
\usepackage{amsmath}
\usepackage{amsfonts}
\usepackage{amssymb}
\usepackage[utf8]{inputenc}
\usepackage[T1]{fontenc}
\usepackage{mathtools}
\usepackage{wasysym}

\DeclareMathAlphabet{\mathds}{U}{BOONDOX-ds}{m}{n}
\usepackage[dvipsnames]{xcolor}

\usepackage{hyperref}
\usepackage{amsthm}
\theoremstyle{definition}

\theoremstyle{plain}

\newcommand{\dd}{\mathrm{d}}

\allowdisplaybreaks

\@ifundefined{textcolor}{}{%
 \definecolor{BLACK}{gray}{0}
 \definecolor{WHITE}{gray}{1}
 \definecolor{RED}{rgb}{1,0,0}
 \definecolor{GREEN}{rgb}{0,1,0}
 \definecolor{BLUE}{rgb}{0,0,1}
 \definecolor{CYAN}{cmyk}{1,0,0,0}
 \definecolor{MAGENTA}{cmyk}{0,1,0,0}
 \definecolor{YELLOW}{cmyk}{0,0,1,0}
 }

\begin{document}
\title{Stable and self-consistent compact star models in teleparallel gravity}

\author{G.G.L. Nashed}%
\email{nashed@bue.edu.eg}
\affiliation{Centre for Theoretical Physics, The British University in Egypt, P.O. Box 43, El Sherouk City, Cairo 11837, Egypt}
\affiliation{Egyptian Relativity Group (ERG), Cairo University, Giza 12613, Egypt}
\author{S. Capozziello}%
\email{capozziello@na.infn.it}
\affiliation{Dipartimento di Fisica ``E. Pancini``, Universit\'a di Napoli ``Federico II'',
Complesso Universitario di Monte Sant' Angelo, Edificio G, Via Cinthia, I-80126, Napoli, Italy}
\affiliation{ Istituto Nazionale di Fisica Nucleare (INFN),  Sezione di Napoli,
Complesso Universitario di Monte Sant'Angelo, Edificio G, Via Cinthia, I-80126, Napoli, Italy}
\affiliation{Laboratory for Theoretical Cosmology, Tomsk State University of Control Systems and Radioelectronics (TUSUR), 634050 Tomsk, Russia.}

\date{\today}
\begin{abstract}
In the framework of Teleparallel Gravity,  we derive a  charged non-vacuum solution   for  a physically symmetric tetrad field with two unknown functions of radial coordinate.  The field equations  result  in a closed-form   adopting   particular  metric potentials and  a suitable  anisotropy function combined with the charge. Under these circumstances, it is possible to obtain  a  set of configurations  compatible with  observed pulsars.  Specifically,  boundary conditions  for the interior spacetime  are applied to the exterior  Reissner-Nordstr\"om metric
 to  constrain   the radial pressure that has to  vanish through
the boundary. Starting from these considerations, we are able  to fix the model parameters.
The pulsar $\textit {PSR J 1614--2230}$, with  estimated
mass $M= 1.97 \pm 0.04\, M_{\circledcirc},$ and  radius $R= 9.69 \pm 0.2$ km
is used  to test  numerically the model. The stability  is studied, through the causality conditions and  adiabatic index, adopting   the  Tolman-Oppenheimer-Volkov
equation.  The mass-radius $(M,R)$ relation  is derived. Furthermore,  the  compatibility of the model  with other  observed pulsars is also studied. We  reasonably conclude that the model can represent realistic compact objects.

\keywords{Teleparallel gravity;  compact stars;
 stellar structure; observational data.} \pacs{11.30.-j; 04.50.Kd; 97.60.Jd.}

\end{abstract}

\maketitle
\section{Introduction}
Soon after the formulation   of General Relativity (GR), several  theories were constructed  in view of fixing as many issues  as possible related to the gravitational field. Among these theories,   there is the one  formulated by H. Weyl which tried  to unify gravitation and  electromagnetism in 1918 \cite{1918SPAW.......465W}. Einstein himself,   in 1928  \cite{doi:10.1002/3527608958.ch37},   adopted  the same philosophy by   Weyl adopting  the  Weitzenb\"ock geometry. In this formulation, one has  to introduce   tetrad fields to describe  dynamics, unlike GR whose dynamical variable is the metric. The tetrad field has 16 components which made Einstein  think that the extra 6 components, with respect to  the   metric, could describe  the components of the electromagnetic field. Nevertheless,  it was shown that these extra 6 components are linked to  the local Lorentz invariance of the theory \cite{MuellerHoissen:1983vc, cirilo}.

Despite the failure of the Weyl and Einstein attempts,  the  new approaches   supplied  the notion of gauge theory and thus the search  for a gauge  gravity started \cite{ORaifeartaigh:1997dvq,Blagojevic:2013xpa}.
In 1979, K. Hayashi and T. Shirafuji \cite{PhysRevD.19.3524} proposed a gravitational theory, called  ``New General Relativity'' that is a gauge theory for  the translation group. This theory involves three free parameters to be determined by the experiment.

Another theory that is built up through the  Weitzenb\"ock geometry is the Teleparallel Equivalent of General Relativity (TEGR).  TEGR and GR are equivalent at the level of  field equations  however, at the level of actions, they are different for  a total divergence  term \cite{Maluf_2013,Wu_2012,Capriolo1,Awad:2017ign,Awad:2017tyz,Hanafy:2015yya}.
 In TEGR, gravity is encoded in  the torsion field, with  vanishing curvature, unlike GR where gravity is encoded in  metric  and curvature fields with vanishing torsion \cite{Li:2010cg,Shirafuji:1997wy,El_Hanafy_2016,Nashed_2011,Krssak:2015lba,Nashed:2003ee,Awad_2018,Bahamonde:2015zma, Cai}.

In this situation, a good approach could be  testing different formulations of  gravitational interaction by  searching for signatures  discriminating among concurrent theories. Systems, in strong field regime, are natural candidates to this aim.  Signatures can come from compact objects like neutron stars or black holes, or from  polarizations of gravitational waves  \cite{Bogdanos,Abedi,Capriolo1,Capriolo2}.

 From a physical point of view, compact objects are, in general,   stars  exhausting
their  nuclear fuel. They can give rise to stable compact objects or to black holes. For example, neutron stars are compact objects which are boosted   by their neutron degeneracy pressure against the attraction of gravity. Another type of compact stars  are the white dwarfs, boosted  by the electron degeneracy pressure against the gravity.

From a theoretical viewpoint, it is well know that the first exact vacuum solution of GR was the Schwarzschild one \cite{Schwarzschild:1916uq}.  Thereafter many solutions investigating   compact stars have been proposed. In particular, searching  for interior  solutions,  describing  realistic compact  objects, became a fertile   domain  to probe   GR and other theories. Using the equation of state (EoS),  one can study the stability structure  of  compact stellar objects and then guess on their internal composition. Specifically,  the  EoS is useful to  investigate the  physical behavior of the stellar structure through the     Tolman-Oppenheimer-Volkov (TOV) equation which is the general relativistic equation of the stellar interior coming from the internal Schwarzschild solution.

In order to  study stellar configurations,  one can assumes, at the beginning,  that the distribution of matter is isotropic which means that the radial and tangential pressures are equal. However, in realistic cases, such assumption does not hold and one  finds that the two components of radial and tangential pressures are not equal: the differences between them create the anisotropy. Such stellar configurations have unequal radial and tangential pressures. Lema\^itre, in 1933,  was the first  which proposed  anisotropic models \cite{Lemaitre:1933gd}. Moreover, it possible to show that, in order   to reach stable configurations, at the maximal  star surface,  the radial pressure has to decrease  and, at the center, it has   to  vanish \cite{PhysRevD.70.024010,Mak:2003kw}.

There are several factors to be taken into account to study   anisotropic stars; among these, the high density regime where the nuclear interactions have to be relativistically treated \cite{Ruderman:1972aj,1975ARA&A..13..335C}. Moreover the existence of a solid core or a 3A type  superfluid may cause the star to be anisotropic \cite{Kippenhahn:1493272}. There is another source that makes the star anisotropic: it is a  strong magnetic field \cite{article3,Astashenok4}. The slow rotation can be considered another   source of  anisotropy \cite{Herrera:1995pm}. Letelier showed that combinations between perfect and null fluids can give rise to  anisotropic fluids \cite{PhysRevD.22.807}.
Other reasons can be taken into account to generate  anisotropies like pion condensation \cite{PhysRevLett.29.382}, strong electromagnetic field \cite{PhysRevD.70.067301} and phase transition \cite{1980JETP...52..575S}.

Dev and Gleiser \cite{Dev2003,Dev2002} and Gleiser and Dev \cite{GLEISER_2004}  investigated the  operators that induce the pressure to be anisotropic. It was shown that the effect of  shear, electromagnetic field, etc. on self-bound systems can be neglected,   if the system
is   anisotropic \cite{Ivanov_2010}. Systems that consist of scalar fields like  boson stars possess anisotropy \cite{Schunck:2003kk}. Gravastars and wormholes can be  considered as anisotropic models \cite{Morris:1988cz,Cattoen:2005he,DeBenedictis_2006, DeFalco}. An application of anisotropic model to stable configurations of neutron stars has been discussed in \cite{Bowers:1974tgi}. They showed  that  anisotropy might have non-negligible effects on  the equilibrium mass and on the surface red-shift.

 A nice study describing the origin and the effects of anisotropy can be found in \cite{Chan:2002bn,10.1093/mnras/287.1.161}. Super dense and  anisotropic neutron stars have been considered and a conclusion is that there is no limiting mass of such stars  \cite{1975A&A....38...51H}.

 Supermassive neutron stars in alternative gravity are considered, for example,  in \cite{Astashenok1,Astashenok2, Astashenok3,Farinelli}. The presence of torsion field in neutron star models is studied in \cite{Feola}. The issue of stability of anisotropic stars has been analyzed in \cite{etde_7103699} and a conclusion is   that the  stability of such systems is  similar to that of  isotropic stars.

 There are many anisotropic models that deal with the anisotropic pressure in the energy-momentum  tensor. Several exact spherically symmetric solutions of interior stellar  have been developed \cite{PhysRevD.26.1262,1984CaJPh..62..239K,1993MNRAS.262.1088B,1999MNRAS.302..337B,1993Ap&SS.201..191B,Barreto2007,Coley_1994,1994MNRAS.271..463M,Singh1995,
Hern_ndez_1999,Harko20,
1995AuJPh..48..635P,PhysRevLett.92.051101,Bohmer2006,Boehmer2007az,Esculpi2007,Khadekar7,Karmakar2007,Abreu2007,
IVANOV2010,PhysRevD.79.087505,Mak2003,doi10.1142/S0217732302008149,Maharaj1989,Chaisi2006,PhysRevD.77.027502,Chaisi2005,Gokhroo1994,PhysRevD.80.064039,
Thomas12,2007IJMPD..16.1479T,Tikekar2005,Thirukkanesh2008,Finch1989,Sharma:2013lqa,2015Ap&SS.356..285P,Bhar2015}.

Beside this, adding   charge effects in compact objects is an issue widely considered in literature, especially for neutron  and quark stars \cite{PhysRevD.4.2185,1985Ap&SS.111..207D,PhysRevD.68.084004,PhysRevD.72.104017,PhysRevD.75.104010,PhysRevD.80.083006}.  An important result is  the fact that the distribution of net charge can improve the maximum mass of compact stars. For example for  white dwarfs, there are several  results  related  to the charge effects affecting the structure (see, e.g.  \cite{PhysRevD.13.2204}).  For this reason, we consider  here  charged compact star models   showing that the net charge effects  can increase the maximum mass.

The aim of the present paper is to derive a novel charged anisotropic solution  in the framework of TEGR  and compare it  with  realistic  stellar configurations using physical assumptions on the form of  metric potential and the  combination of charge and anisotropy. In particular, we consider   the pulsar $\textit {PSR J 1614 - 2230}$ which   estimated
mass    $M= 1.97 \pm 0.04 M_{\circledcirc}$ and radius is  $R = 9.69 \pm 0.2$ km  \cite{Demorest}. This peculiar system  escapes the standard GR explanation of neutron stars because it is too massive to be stable unless   one assumes exotic EoS or alternative gravities.
 We discuss  the physical parameters of such a pulsar considering our solution  derived in the framework of TEGR. From our point of view, this can constitute a possible  test for the theory.

The set up of the paper  is  the following. In Sec. \ref{S2},  we sketch the TEGR theory in presence of the electromagnetic field. Sec. \ref{S3} is devoted to the discussion of charged compact stars in TEGR.
The requirements for a physically consistent stellar model in TEGR are discussed in Sec. \ref{S4}.
The physical properties of the model are considered in Sec.\ref{property}. The model is matched with realistic compact stars, in particular with  $\textit {PSR J 1614--2230}$, in    Sec. \ref{data}.  In Sec.\ref{stability}, the stability of the model is taken into account with respect to the TOV equation and the adiabatic index. In particular, we report the $(M,R)$ relation and how it is affected by the electromagnetic field.  Discussion and conclusions are reported in Sec. \ref{S5}.

\section{Teleparallel equivalent of general relativity and the electromagnetic field}\label{S2}

In  TEGR theory, at each point of the spacetime manifold, $M$, can be defined  a tangent  Minkowski spacetime $\eta_{i j}=(+1,-1,-1,-1)$\footnote{ Latin indices represent spacetime coordinates. Greek ones describe tangent space indices.}. As is well-known that the dynamical fields of TEGR   are the four linearly independent vierbeins (tetrads) from where we can define the metric and its inverse as
\begin{align}  \label{metr}
	g_{\mu\nu} = \eta_{ab}l^a{}_\mu l^b{}_\nu \,, \qquad \qquad g^{\mu\nu} = \eta^{ab}L_a{}^\mu L_b{}^\nu \,,
\end{align}
with $ \eta_{ab}$  being the flat Minkowski metric of the tangent space and $l^b{}_\nu$, $L_b{}^\nu$
 are the covariant and
 contra-variant  tetrad fields. Using Eq. (\ref{metr}) we can define the following orthogonal conditions
\begin{align} L_i{}^\mu l^i{}_\nu &= \delta^\nu_\mu,   &  l^i{}_\mu L_j{}^\mu &= \delta^i_j\,.\end{align}
From the above definitions,  we can define the torsion tensor as\footnote{Square brackets means anti-symmetrization, i.e., $A_{[\mu \nu]}=\frac{1}{2}(A_{[\mu \nu]}-A_{[\nu \mu]})$.}
\begin{align}  \label{tor}
	T^a{}_{\mu\nu} = 2 \left(\partial_{[\mu}l^a{}_{\nu]} + \omega^a{}_{b[\mu} l^b{}_{\nu]}\right)\,.
\end{align}
where $\omega^a{}_{b\mu} $ is the spin connection which we can set equal zero due to  TEGR.  Therefore the torsion tensor (\ref{tor})  reduces to \begin{align}
T^a{}_{\mu\nu} = \partial_{[\mu}l^a{}_{\nu]}.\end{align}
TEGR  with electromagnetic field is recovered from the action
\begin{align}
	S_{\rm TEGR}= \int \dd^4x\ |l| \left( \frac{1}{2\kappa^2}  T + \mathcal{{\cal F}} \right)\,,\label{TEGR}
\end{align}
where $\kappa^2=8\pi$, $|l|=\det (l^a{}_\mu)=\sqrt{-g}$ is the determinant of the tetrad field,  ${\cal {\cal F}}$ is  a gauge-invariant  scalar defined as ${\cal F} = \frac{1}{4}{\cal F}_{\alpha \beta}{\cal F}^{\alpha \beta}$ \cite{plebanski1970lectures,Nashed:2004pn}.  The torsion scalar $T$ is defined  as
\begin{align} \label{tors}
	T =  T^a{}_{\mu\nu}S_a{}^{\mu\nu} =\frac{1}{2} \left(L_a{}^\sigma g^{\rho \mu} L_b{}^\nu + 2 L_b{}^\rho g^{\sigma \mu} L_a{}^\nu + \frac{1}{2} \eta_{ab} g^{\mu\rho} g ^{\nu\sigma} \right) T^a{}_{\mu\nu} T^b{}_{\rho\sigma}\,,
\end{align}
where $S_a{}^{\mu\nu}$ is the superpotential  defined as

\begin{equation}
S_a{}^{\mu\nu} = \frac{1}{2}(K^{\mu\nu}{}_a - L_a{}^\mu T_\lambda{}^{\lambda\nu} + L_a{}^\nu T_\lambda{}^{\lambda\mu})
\end{equation}
in terms of the contortion tensor
\begin{equation}
K^{\mu\nu}{}_a = \frac{1}{2}(T^{\nu\mu}{}_a + T_a{}^{\mu\nu} - T^{\mu\nu}{}_a)\,.
\end{equation}
 The  metric components are   functions of the tetrads.
The variation with respect to the tetrad $l^a{}_\mu$ yields the field equations (see \cite{Krssak:2015oua} for details)
\begin{align}\label{eq:fT}
	\frac{1}{4}T L_a{}^\mu +  T^b{}_{\nu a} S_b{}^{\mu \nu } + \frac{1}{h}\partial_{\nu}(l S_a{}^{\mu \nu })   &= \frac{1}{2}\kappa^2 {\cal T}_a{}^\mu\,.
\end{align}
The stress-energy tensor $ {\cal T}_a{}^\mu$   consists of two terms
 \begin{equation} \label{max}
{\cal T}_a{}^\mu\:={\cal T}_a{}^\mu{}_{AF}+{\cal T}_a{}^\mu_{EM}, \end{equation}
where
\begin{equation}
{\cal T}_a{}^\mu{}_{AF}=(p_t+\rho)u^\mu u_a+p_t\delta_a{}^\mu+(p_r-p_t)\xi_a \xi^\mu\,,
\end{equation}
 is the energy-momentum  tensor  of an anisotropic fluid, and
\begin{equation}
{\cal T}_a{}^\mu{}_{EM}=\frac{1}{4}\left({\cal F}_{a \alpha} {\cal F}^{\mu \alpha}-\frac{1}{4} \delta_a{}^\mu {\cal F}_{\alpha \beta} {\cal F}^{\alpha \beta}\right)\,,
\end{equation}
 is the energy-momentum tensor of the electromagnetic field.
Here $u_\mu$ is the time-like vector defined as $u^\mu=[1,0,0,0]$ and $\xi_\mu$ is the unit space-like vector in the radial direction defined as $\xi^\mu=[0,1,0,0]$ such that $u^\mu u_\mu=-1$ and $\xi^\mu\xi_\mu=1$. In this paper, $\rho$ is the energy-density, $p_r$  and $p_t$ are the radial and  the tangential pressures respectively. Furthermore, the electromagnetic
tensor $F_{\mu \nu}$  satisfies the Maxwell equations
\begin{eqnarray}\label{max1}
&&F_{\mu \nu,\gamma}+F_{\nu \gamma,\mu}+F_{\gamma \mu, \nu}=0,\nonumber\\
&&\left(\sqrt{-g}F^{\mu \nu}\right)_{\, ,\nu}=4\pi J^\mu \sqrt{-g},
\end{eqnarray}
where $J^\mu=\sigma u^\mu$ is the  current density and $\sigma$ is the charge density. In the next section, we are going to apply the field equations (\ref{eq:fT}) and (\ref{max1}) to a spherically symmetric tetrad space and try to solve the resulting system of differential equations.
\section{Charged compact stars  }\label{S3}
 Let us begin  with the following spherically symmetric metric using the spherical coordinates $(t,r,\theta, \phi)$
\begin{equation}
ds^2=-G(r) \,dt^2+H(r)dr^2+r^2(d\theta^2+\sin^2\theta d\phi^2)\,,\label{met1}
\end{equation}
where $G(r)$ and $H(r)$ are two  unknown functions depending on the radial coordinate $r$. The above metric (\ref{met1}) can be reproduced from the following covariant tetrad field ~\cite{Bahamonde:2019zea}

\begin{equation}
l^a{}_{\mu}=\left(
\begin{array}{cccc}
\sqrt{G(r)} & 0 & 0 & 0 \\
0 & \sqrt{H(r)} \cos (\phi ) \sin (\theta ) & r \cos (\phi ) \cos (\theta )  & -r \sin (\phi ) \sin (\theta )  \\
0 & \sqrt{H(r)} \sin (\phi ) \sin (\theta )  & r \sin (\phi ) \cos (\theta )  & r \cos (\phi ) \sin (\theta ) \\
0 & \sqrt{H(r)} \cos (\theta ) & -r \sin (\theta ) & 0 \\
\end{array}
\right)\label{tet}\,.
\end{equation}
 The tetrad (\ref{tet}) is the  output  of the product of a diagonal tetrad and a local Lorentz transformation which can be written as

\begin{eqnarray} \label{tetdcom}
&& l^a{}_{\mu}=\Lambda^a{}_b l^b{}_{\mu_{diag}}\nonumber\\
&&\Rightarrow \left(
\begin{array}{cccc}
\sqrt{G(r)} & 0 & 0 & 0 \\
0 & \sqrt{H(r)} \cos (\phi ) \sin (\theta ) & r \cos (\phi ) \cos (\theta )  & -r \sin (\phi ) \sin (\theta )  \\
0 & \sqrt{H(r)}\sin (\phi ) \sin (\theta )  & r \sin (\phi ) \cos (\theta )  & r \cos (\phi ) \sin (\theta ) \\
0 & \sqrt{H(r)} \cos (\theta ) & -r \sin (\theta ) & 0 \\
\end{array}
\right)\nonumber\\
&&\equiv\underbrace{\left(
\begin{array}{cccc}
1 & 0 & 0 & 0 \\
0 & \cos (\phi ) \sin (\theta ) &  \cos (\phi ) \cos (\theta )  & - \sin (\phi )   \\
0 &  \sin (\phi ) \sin (\theta )  &  \sin (\phi ) \cos (\theta )  &  \cos (\phi )  \\
0 &\cos (\theta ) & -\sin(\theta) & 0 \\
\end{array}
\right)}_{Local\, \, Lorentz \, \,transformation}\times\underbrace{\left(
\begin{array}{cccc}
\sqrt{G(r)} & 0 & 0 & 0 \\
0 & \sqrt{H(r)} &  0  & 0   \\
0 & 0  &  r  & 0  \\
0 &0 & 0 & r \sin(\theta) \\
\end{array}
\right)}_{Diagonal\, \, tetrad} \label{tet1}\,.\end{eqnarray}

It is worth  noticing that the diagonal tetrad can be  applied to the field equations of $f(T)$ TEGR models \cite{Bahamonde:2019zea}.  Here we shall take into account the effect of local Lorentz transformation given by Eq. (\ref{tet1}) and see if the results are different from those presented in \cite{Singh:2019ykp}.

Using Eq. (\ref{tet}) into Eq. (\ref{tors}),  the torsion scalar takes the form
\begin{align}\label{torsc}
  T= \frac{2\left(2G\sqrt{H}-GH+r\sqrt{H}G'-G-rG'\right)}{GHr^2}\,.
\end{align}
Despite the fact  that the diagonal tetrad   in Eq. (\ref{tetdcom}) and  tetrad (\ref{tet}) reproduce the same metric, their  torsion scalars are different. The  reason for this difference is the Local Lorentz transformation (LLT) in (\ref{tetdcom}). The torsion scalar of Eq. (\ref{torsc}) goes to zero when  both  the unknown functions $G(r)$ and $H(r)$ tend to one, which is the necessary condition to achieve the asymptotic   flatness.  Nevertheless, the torsion scalar presented in \cite{Singh:2019ykp} does not vanish as it should when we apply the same condition. So the existence of LLT makes our  torsion scalar more physically motivated.

Using Eq. (\ref{torsc}) into   field Eqs. (\ref{eq:fT}) and  (\ref{max1}),   we get
\begin{eqnarray} \label{syd}
 -8\pi \rho+E^2(r)&=& \frac{H[1-H]-rH'}{r^2H^2}\,,\nonumber\\
8\pi p_r+E^2(r)&=&\frac{rG'-G(H-1)}{2r^2GH}\,,\nonumber\\
8\pi p_t-E^2(r)&=&\frac{2GH[G'+rG'']-rHG'^2-2G^2H'-rGG'H'}{8rG^2H^2}\,,\\
\Delta(r)&=&\frac{r^2H[2GG''-G'^2]-rGG'[2H+rH']-2G^2[2H(1-H)+rH']}{8r^2G^2H^2}+2E^2(r)\,,\nonumber\\
\sigma(r)&=&\frac{\sqrt{H(r)}(r^2E)'}{4\pi r^2},\label{fec}\nonumber
\end{eqnarray}
where the prime  indicates derivatives w.r.t the radial coordinate $r$ and $\Delta=8\pi(p_t-p_r)$ is the pressure difference.  We  used the energy-momentum tensor ${\cal T}^{\mu}_{\nu}$ to get the form
\begin{align}\label{enmo}
  \mbox{diag}\,{\cal T}^\mu_{\nu}= \left(-\rho+E^2(r)\,,p_r+E^2(r)\,,p_t-E^2(r)\,,p_t-E^2(r)\right)\,,
\end{align}
where $E^2(r)$ is the electric field defined as
\begin{align}\label{elct}
E(r) = \frac{q(r)}{r^2}\, , \qquad \textrm{and\, q(r)\, is \, defined \, as}\qquad \qquad q(r) =4\pi\int_{0}^r \sigma(\xi) \xi^2d\xi\,,
\end{align}
with $q(r)$ being the electric charge inside a sphere of radius $r$ and $\sigma$ is the charge
density (for details see, e.g.,\cite{Cooperstock1978,article1}). Here $\Delta(r)$ is the anisotropic
parameter of the stellar system.  Field Eqs. (\ref{syd})   coincide  with  those given in \cite{Singh:2019ykp},  and can be recast also in  the field equations  presented in \cite{doi:10.1098/rspa.1974.0065}  and in \cite{Das_2003}.

 The above differential system are  five independent equations in seven unknown  functions,  $G$, $H$, $\rho$, $p_r$, $p_t$, $E(r)$ and $\Delta(r)$. Therefore, we need two extra conditions  to solve the above system. One of these extra conditions is   assuming  the metric potential $g_{rr}$  having the form
 \begin{align}\label{pot}
H(r) =\frac{1}{(1-\frac{r^2}{k^2})^4}\,,
\end{align}
where $k$ is a constant having the  dimension of the length.
It will be determined from the matching conditions. Eq. (\ref{pot}) shows that, for  $r=0$ and $H(r)=1$,   $H(r)$ is finite at the center. Also the derivative of $H$ is finite at the origin. The second condition comes from the use of Eq. (\ref{pot}) in  $\Delta(r)$ which takes the form
\begin{eqnarray}\label{d1}
&& \Delta(r)=\frac{24r^3k^4-32r^5k^2+12r^7+16k^8k_1{}^2 r^3}{8rk^8} \nonumber\\
&&+\frac{1}{8rk^8}\left[r\left(2k^8+12r^4k^4-8r^6k^2+2r^8-8r^2k^6\right)\frac{G''}{G}-\left(2k^8-12r^4k^4+16r^6k^2-6r^8\right)\frac{G'}{G}\right] \nonumber\\
&&+\left(\frac{4r^7k^2-rk^8+4r^3k^6-6r^5k^4-r^9}{8rk^8}\right)\frac{G'^2}{G^2},
\end{eqnarray}
where we have put $E^2(r)=k_1^2r^2$.  Eq.  (\ref{d1}) is a second order differential equation in the unknown $G$  function. In order  to solve  Eq.  (\ref{d1}),  we assume the anisotropic $\Delta$ to have the form
\begin{eqnarray}\label{d2}
 \Delta(r)=\frac{24r^2k^4-32r^4k^2+12r^6+16k^8k_1{}^2 r^2}{8k^8}.
\end{eqnarray} 
Eq. (\ref{d2}) has not been used before for charged stars in GR and TEGR. The present aim  is to study the effect of  anisotropy, given in  (\ref{d2}), in  presence of  electric charge, on the  neutron star structure. Inserting Eq. (\ref{d2}) into Eq. (\ref{d1}),  we get
\begin{eqnarray}\label{d1r}
&& \frac{1}{8rk^8}\left[r\left(2k^8+12r^4k^4-8r^6k^2+2r^8-8r^2k^6\right)\frac{G''}{G}-\left(2k^8-12r^4k^4+16r^6k^2-6r^8\right)\frac{G'}{G}\right] \nonumber\\
&& +\left(\frac{4r^7k^2-rk^8+4r^3k^6-6r^5k^4-r^9}{8rk^8}\right)\frac{G'^2}{G^2}=0\,,
\end{eqnarray}
 The solution of  differential equation (\ref{d1r}) is
{\begin{equation}\label{Gs}
 G=\frac{c_1(1+c_2r^2)^2}{(r^2-k^2)^2}\,.
\end{equation}
Clearly, $G$ goes to a constant value for $r\rightarrow 0$ and $r\rightarrow \infty$.

From Eqs. (\ref{pot}) and (\ref{Gs}) of the system of differential Eqs. (\ref{syd}),  we get the remaining  unknown functions 
\begin{eqnarray}\label{sol}
&&\rho=\frac{12k^6-30k^4r^2+28k^2r^4-9r^6+2k^8k_1{}^2r^2}{8\pi k^8}\,, \nonumber\\
 &&p_r=\frac{c_2r^8-r^6(3+8k^2c_2)+2r^4(9k^4c_2-k^8c_2k_1{}^2+4k^2)-2r^2(3k^4+k^8k_1{}^2+8k^6c_2)+4k^8c_2}{8\pi k^8(1+c_2r^2)^2}\,,  \\
 &&p_t=\frac{c_2[k^8(2+r^4k_1{}^2)-8r^2k^6+12r^4k^4-8k^2r^6+2r^8]+k^8r^2k_1{}^2}{4\pi k^8(1+c_2r^2)^2}\,, \qquad \sigma(r)=\frac{3k_1(k^2-r^2)^2}{4\pi k^4}\,.\nonumber
\end{eqnarray}
 If the constant $k_1=0$,  our star is neutral. In this case,  Eqs. (\ref{sol}) will be identical with those presented in  \cite{Singh:2019ykp}.  Eqs (\ref{sol}), at the boundary of the star, take the form
\begin{eqnarray}\label{solb}
&&\rho=\frac{12k^6-30k^4R^2+28k^2R^4-9R^6+2k^8k_1{}^2R^2}{8\pi k^8}\,, \nonumber\\
 &&p_r=\frac{c_2R^8-R^6(3+8k^2c_2)+2R^4(9k^4c_2-k^8c_2k_1{}^2+4k^2)-2R^2(3k^4+k^8k_1{}^2+8k^6c_2)+4k^8c_2}{8\pi k^8(1+c_2R^2)^2}\,,  \\
 &&p_t=\frac{c_2[k^8(2+r^4k_1{}^2)-8R^2k^6+12R^4k^4-8k^2R^6+2r^8]+k^8r^2k_1{}^2}{4\pi k^8(1+c_2R^2)^2}\, ,\nonumber\\
\end{eqnarray}
 where $R$ is the radius at the boundary of the star. From the second equation of (\ref{solb}), we  get the condition for the  vanishing of radial pressure, that is
\begin{align}\label{vanr}
c_2=\frac{R^2(3R^4-8R^2k^2+6k^4+2k^8k_1{}^2)}{R^8-8R^6k^2-18R^4k_1{}^2+2R^4k^8k_1{}^2+16R^2k^6-4k^8}\,.
\end{align}
For $R=8.51M_{\odot}$, which is the boundary radius of the star $\it{Her X-1}$, $c_2$ takes the form
\begin{align}\label{vanr1}
c_2=\frac{3-8r^2k^2+6k^4+2k^8k_1{}^2}{1-8k^2-18k_1{}^2+2k^8k_1{}^2+16k^6-4k^8}\,.
\end{align}
If we substitute $k=33.43508140$, $k1=0.008680552960$ in Eq. (\ref{vanr1}),  we get $c_2=0.005$. In other words,  if we substitute $R=8.51M_{\odot}$, $k=33.43508140$ and $k1=0.008680552960$ in the second equation of (\ref{solb}), we get a vanishing radial pressure for the  pulsar $\textit {Her X-1}$.
It is important to mention that the anisotropic force is defined as $\frac{2\Delta}{r}$ and it is attractive for $p_r-p_t>0$ and  repulsive
 for $p_r-p_t<0$. The mass contained within a radius $r$ of the sphere
is defined as
\begin{align}\label{mas}
M(r)={\int_0}^r [4\pi \rho(\xi) +E(\xi)]\xi^2 d\xi\,.\end{align}
Using Eq. (\ref{sol}) in Eq. (\ref{mas}),  we get
\begin{align}\label{mas1}
M(r)=\frac{r^3(20k^6+20k^2r^4-5r^6-30k^4r^2+2k^8k_1{}^2 r^2)}{10k^8}\,.\end{align}
The compactness parameter of a spherically symmetric
source with radius $r$  takes the form \cite{Singh:2019ykp}
\begin{eqnarray}\label{gm1}
&&u(r)=\frac{2M(r)}{r}=\frac{r^2(20k^6+20k^2r^4-5r^6-30k^4r^2+2k^8k_1{}^2 r^2)}{5k^8}.
\end{eqnarray}

In the next section, we are going to discuss  the physical requirements to derive  viable stellar structures and to see if model  (\ref{sol}) satisfy them or not.
\section{Requirements for a physically consistent stellar model}\label{S4}
A physically viable stellar model has to  satisfy the following
conditions throughout the stellar configurations:\vspace{0.1cm}\\
$\bullet$ The gravitational potentials $G(r )$ and $H(r )$,  and the matter
quantities $\rho$, $p_r$ , $p_t$ have to  be well defined at the
center and regular as well as singularity free throughout
the interior of the star.\vspace{0.1cm}\\
$\bullet$  The energy density $\rho$ has to  be positive throughout
the stellar interior i.e., $\rho\geq 0$. Its value at the center
of the star should be positive, finite and monotonically
decreasing towards the boundary inside the stellar
interior, that is  $\frac{d\rho}{dr}\leq 0$.\vspace{0.1cm}\\
$\bullet$  The radial pressure $p_r$ and the tangential pressure $p_t$
must be positive inside the fluid configuration i.e.,
$p_r\geq0$, $p_t\geq0$. The gradient of the pressure must be
negative inside the stellar body, i.e., $\frac{dp_r}{dr}< 0$, $\frac{dp_t}{dr}< 0$.
At the stellar boundary $r= R$,  the radial pressure $p_r$
has to vanish but the tangential pressure $p_t$ may not be
zero at the boundary.
At the center both pressures are equal. This
means the anisotropy has to vanish at the center, that is
	$\Delta(r = 0) = 0$.\vspace{0.1cm}\\
$\bullet$  For an anisotropic fluid sphere, the  fulfillment of   the  energy conditions refers to the following
inequalities in every point inside the fluid sphere:\vspace{0.1cm}\\
(i) Weak energy condition (WEC): $p_r+\rho > 0$, $\rho> 0$.\vspace{0.1cm}\\
(ii) Null energy condition (NEC): $p_t+\rho > 0$, $\rho> 0$.\vspace{0.1cm}\\
(iii) Strong energy condition (SEC): $p_r+\rho > 0$, $p_t+\rho > 0$, $\rho-p_r-2p_t > 0$.\vspace{0.1cm}\\
(iv) Dominant energy conditions (DEC): $\rho\geq \lvert p_r\lvert$ and
$\rho\geq \lvert p_t\lvert$ .\vspace{0.1cm}\\
$\bullet$  Causality condition has to be satisfied to get a  realistic
model i.e. the speed of sound must be smaller than 1
(assuming the speed of light $c = 1$) in the interior of
the star, i.e. $1\geq\frac{dp_r}{dr}\geq 0$, $1\geq \frac{dp_t}{dr}\geq 0$.\vspace{0.1cm}\\
$\bullet$  The interior metric functions have to  match smoothly
the exterior Schwarzschild metric at the
boundary.\vspace{0.1cm}\\
$\bullet$  For a stable model, the adiabatic index should be
greater than $\frac{4}{3}$.\vspace{0.1cm}\\
$\bullet$  Herrera method \cite{HERRERA1992206}   to study the stability of
anisotropic stars suggests that a viable model should
also satisfy $0>v_r{}^2-v_t{}^2>-1$ where $v_r$ and $v_t$ are
the radial and transverse speed respectively.

Now we are going to analyze  the above physical requirements in details to  see if our model satisfy them.

\section{Physical properties  of the model}\label{property}
Let us  test the model (\ref{sol}) and see if it is consistent  with realistic  stellar structures. To this aim,  we discuss the following issues:
\subsection{Non-singular model}

i- The metric functions of this model satisfy,
\begin{align}\label{sing}
G(0)=\frac{c_1}{k^4}\,\qquad  \textrm{and} \qquad H(0)=1,
\end{align}
which means that the gravitational potentials are finite at the center of the stellar configuration. Moreover, the derivatives of these potentials are finite at the center, i.e., $G'_{r=0}=H'_{r=0}=0$. The above conditions means that the metric is regular at the center and has a well
behavior  throughout the interior of the stellar.\vspace{0.1cm}\\

{ii- Density, radial and tangential pressures of (\ref{sol}), at the center, have the form
\begin{align}\label{reg}
\rho(0)=\frac{3}{4\pi k^2}\, \qquad p_r(0)=p_t(0)=\frac{c_2}{4\pi}.
\end{align}
Eqs. (\ref{reg}) show that the density is always positive and the anisotropy is vanishing at the center. The radial and tangential pressures  have a positive value as soon as  $c_2>0$ otherwise they become negative. Moreover, the Zeldovich condition \cite{1971reas.book.....Z} states that the radial pressure must be less than or equal to the density at the center i.e., $\frac{p_r(0)}{\rho(0)}\leq 1$. Using the Zeldovich condition in Eq. (\ref{reg}), we get
\begin{align}\label{reg1}
c_2k^2\leq 3.
\end{align}

iii- The derivative of energy density, radial and tangential
pressures  of the  model are respectively:
\begin{eqnarray}\label{dsol}
&&\rho'=-\frac{r(30k^4+27r^4-2k^8k_1{}^2-56k^2r^2)}{8\pi k^8}, \qquad \qquad p'_r=-\frac{1}{8\pi k^8(1+c_2r^2)^2}\Big[r\Big(2k^2\{k^2(k^4k_1{}^2(c_2r^2+1)^2 \nonumber\\
 &&+[8k^2c_2+3])+8r^2[c_2r^2(c_2r^2+1)-1]-9k^2c_2r^2[c_2r^2+2]\}-c_2{}^2[3r^9-4k^8]+r^4[9-2c_2r^2]\Big)\Big], \nonumber\\
 &&p'_t=-\frac{r(2c_2k^6[4+k^2c_2]-k^8k_1{}^2(c_2r^2+1)^2-12k^4c_2r^2(c_2r^2+2)+8k^2c_2r^4[2c_2r^2+3]-2c_2r^6[4+3c_2r^2])}{4\pi k^8(1+c_2r^2)^2}\, ,
\end{eqnarray}
where $\rho'=\frac{d\rho}{dr}$, $p'_r=\frac{dp_r}{dr}$ and $p'_t=\frac{dp_t}{dr}$. Eqs. (\ref{dsol}) show that the gradients of density, radial and tangential pressures are negative as we will show when we plot them.\vspace{0.1cm}\\
iv-The radial and transverse velocity of sound (c = 1) are
obtained as
\begin{eqnarray}\label{dso2}
&&v_r{}^2=\frac{dp_r}{d\rho}=\frac{1}{(c_2r^2+1)^2(30k^4+27r^4-2k^8k_1{}^2-56k^2r^2)}\Big(2k^2\{k^2(k^4k_1{}^2(c_2r^2+1)^2 \nonumber\\
 &&+[8k^2c_2+3])+8r^2[c_2r^2(c_2r^2+1)-1]-9k^2c_2r^2[c_2r^2+2]\}-c_2{}^2[3r^9-4k^8]+r^4[9-2c_2r^2]\Big), \nonumber\\
 &&v_r{}^2=\frac{dp_t}{d\rho}=\frac{2(2c_2k^6[4+k^2c_2]-k^8k_1{}^2(c_2r^2+1)^2-12k^4c_2r^2(c_2r^2+2)+8k^2c_2r^4[2c_2r^2+3]-2c_2r^6[4+3c_2r^2])}{(1+c_2r^2)^2
 (30k^4+27r^4-2k^8k_1{}^2-56k^2r^2)}\, .\nonumber\\
 &&
\end{eqnarray}
\subsection{Matching conditions}
We assume that the  exterior spacetime for a not-rotating star is empty and
 described by the exterior Reissner-Nordstr\"om solution that is a solution of vacuum  teleparallel gravity as shown in \cite{Nashed:2007sc,Nashed:2013bfa,Nashed:2006ff}.  It has the form
\begin{eqnarray}\label{Eq1}  ds^2= -\Big(1-\frac{2M}{r}+\frac{Q^2}{r^2}\Big)dt^2+\Big(1-\frac{2M}{r}+\frac{Q^2}{r^2}\Big)^{-1}dr^2+r^2d\Omega^2,
 \end{eqnarray}
where $M$ is the total mass, $Q$ is the charge   and $\frac{Q^2}{(r-2M)}<1$.
We have to  match the interior spacetime metric (\ref{met1}) with
the exterior Reissner-Nordstr\"om spacetime metric  (\ref{Eq1}) at the
boundary of the star $r = R$. The continuity of the metric
functions across the boundary $r = R$ gives the conditions
\begin{eqnarray}\label{Eq2}  G(r=R)=(1-\frac{2M}{R}+\frac{Q^2}{R^2}), \qquad \qquad H(r=R)=(1-\frac{2M}{R}+\frac{Q^2}{R^2})^{-1},
 \end{eqnarray}
in addition to the fact that radial pressure approaches to zero at a finite value of the radial
parameter $r$ which coincides with the radius of the star $R$. Therefore, the radius
of the star can be obtained by using the physical condition $p_r (r =R) = 0$. From the above conditions, we get the constraints on the constants $c_1$, $c_2$ and $k$. The  constant $k$, from these conditions, is
\begin{eqnarray}\label{Eq3} k=\pm\frac{R\sqrt{
R(R\pm
\sqrt{A(R)})\pm(A(R)+R)\sqrt{RA(R)^{1/2}}}}{\sqrt{B(R)}},
 \end{eqnarray}
 where $A(R)=R^2-2MR-Q^2$ and $B(R)=2MR-Q^2$.  The constants $c_1$ and $c_2$ are lengthy and useless to be reported here.  We shall write their numerical values when confronting with observational data.
 \section{Matching the model with realistic compact stars} \label{data}
 Let us consider now the previous  physical conditions of the  model derived  to test it
by using masses and radii of observed pulsars. In order to support  our model, we will study
the pulsar \textrm{PSR J 1614 - 2230} whose estimated mass and
radius are $M = 1.97\pm 0.04 M_\circledcirc$ and $R = 9.69\pm0.2$ km, respectively \cite{Gangopadhyay:2013gha}.
We can use  the maximal values  $M=2.01 M_\circledcirc$ and  $R=9.89$km as  input parameters.
The boundary conditions are adopted  to determine the
constants  $k=28$, $c_1=174908.5108$ and $c_2=0.0003$.
Adopting these  constants, we can plot the  physical quantities.
The regular behavior of these  one can be assumed as a first  requirement to fit
 a realistic star  model.

 Figs. \ref{Fig:1} \subref{fig:pot1} and \ref{Fig:1} \subref{fig:pot2}
represent  the  behavior of metric potentials for
\textrm{PSR J 1614--2230}. As Fig.  \ref{Fig:1} shows,  the metric potentials assume the values  $H(0)=1$ and $G(0)=0.2733207699$ for $r=0$. This means that both of them are finite and positive at the center.
\begin{figure}
\centering
\subfigure[~Metric potential H(r)]{\label{fig:pot1}\includegraphics[scale=0.4]{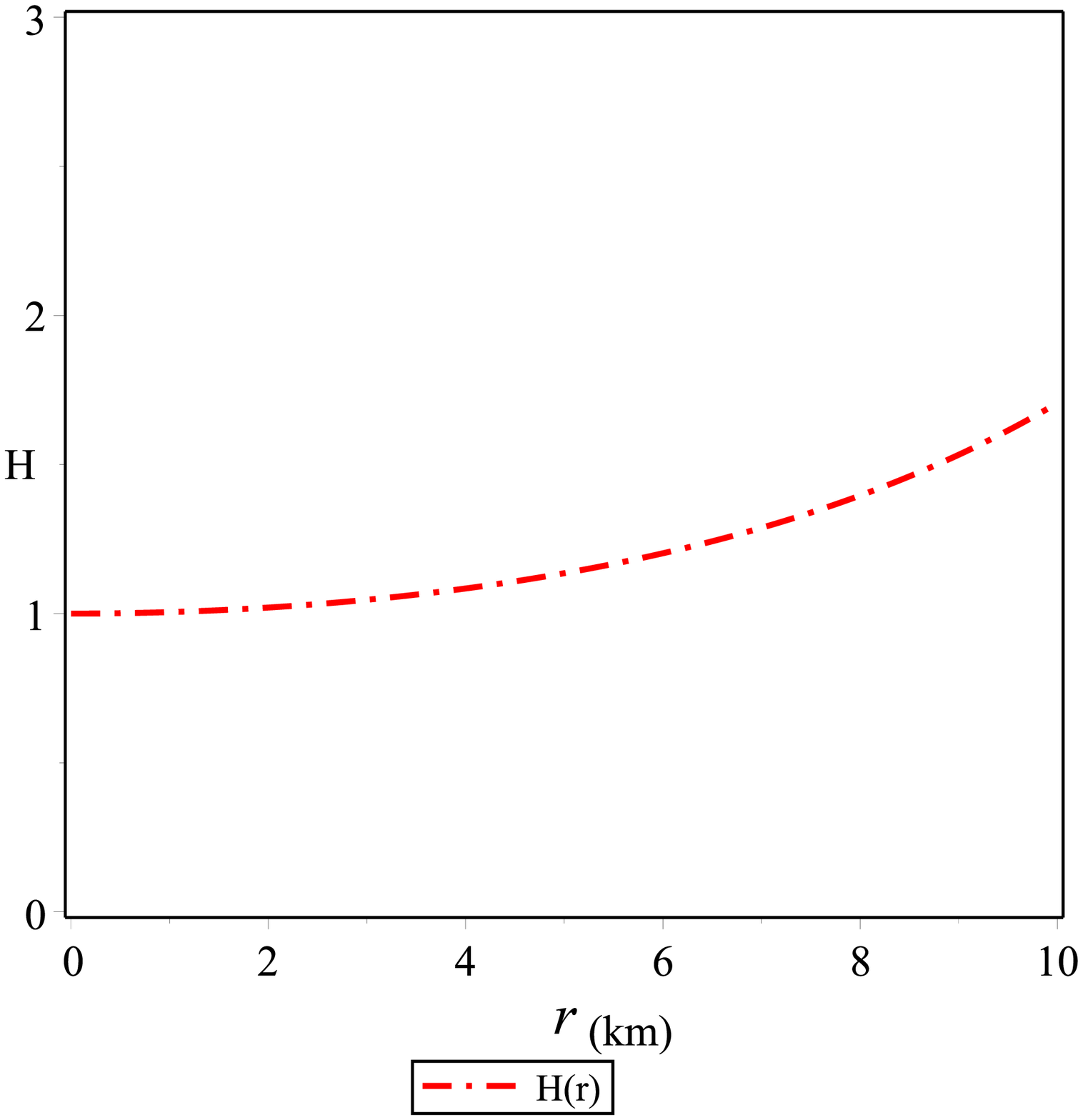}}
\subfigure[~Metric potential G(r)]{\label{fig:pot2}\includegraphics[scale=.4]{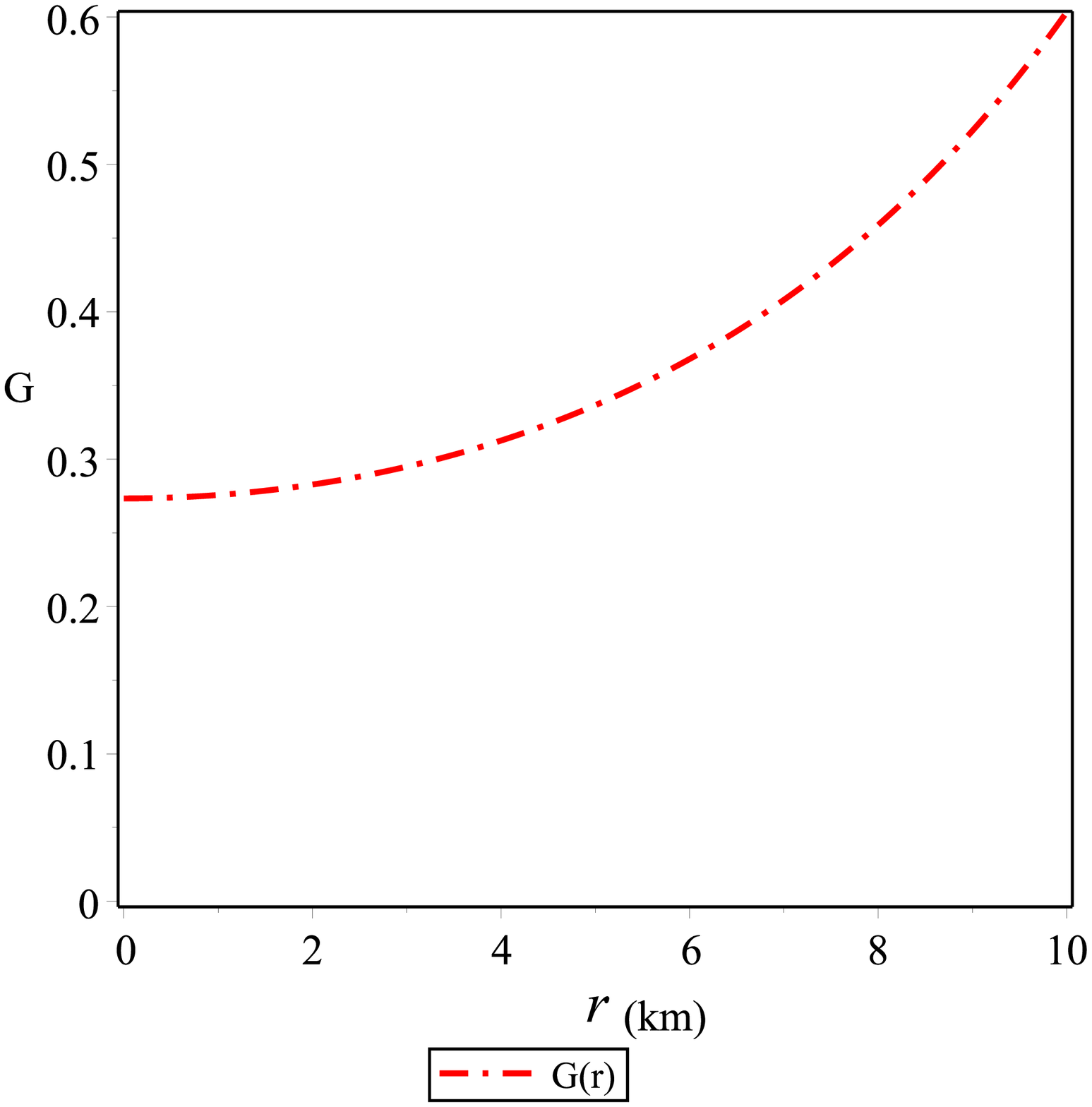}}
\caption[figtopcap]{\small{{Schematic plots of  the metric potentials (\ref{pot}) and  (\ref{Gs}) vs  the radial coordinate  $r$ in Km  using the constants constrained by PSR J 1614--2230 data.}}}
\label{Fig:1}
\end{figure}

Density, radial and tangential pressures,  given in Eq.  (\ref{sol}), are plotted in Fig. \ref{Fig:2}.
\begin{figure}
\centering
\subfigure{\label{fig:density}\includegraphics[scale=0.4]{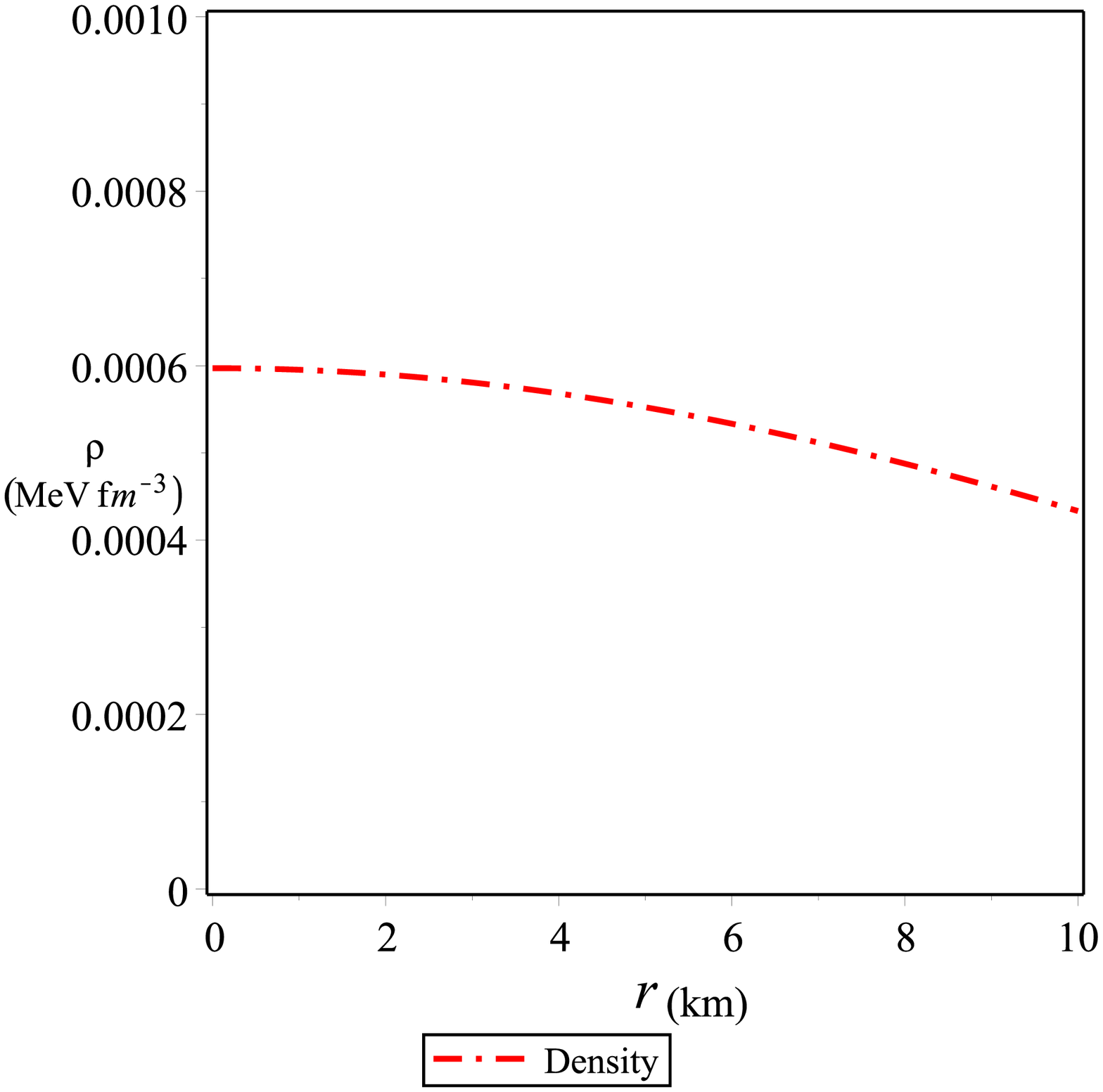}}
\subfigure{\label{fig:pressure}\includegraphics[scale=.4]{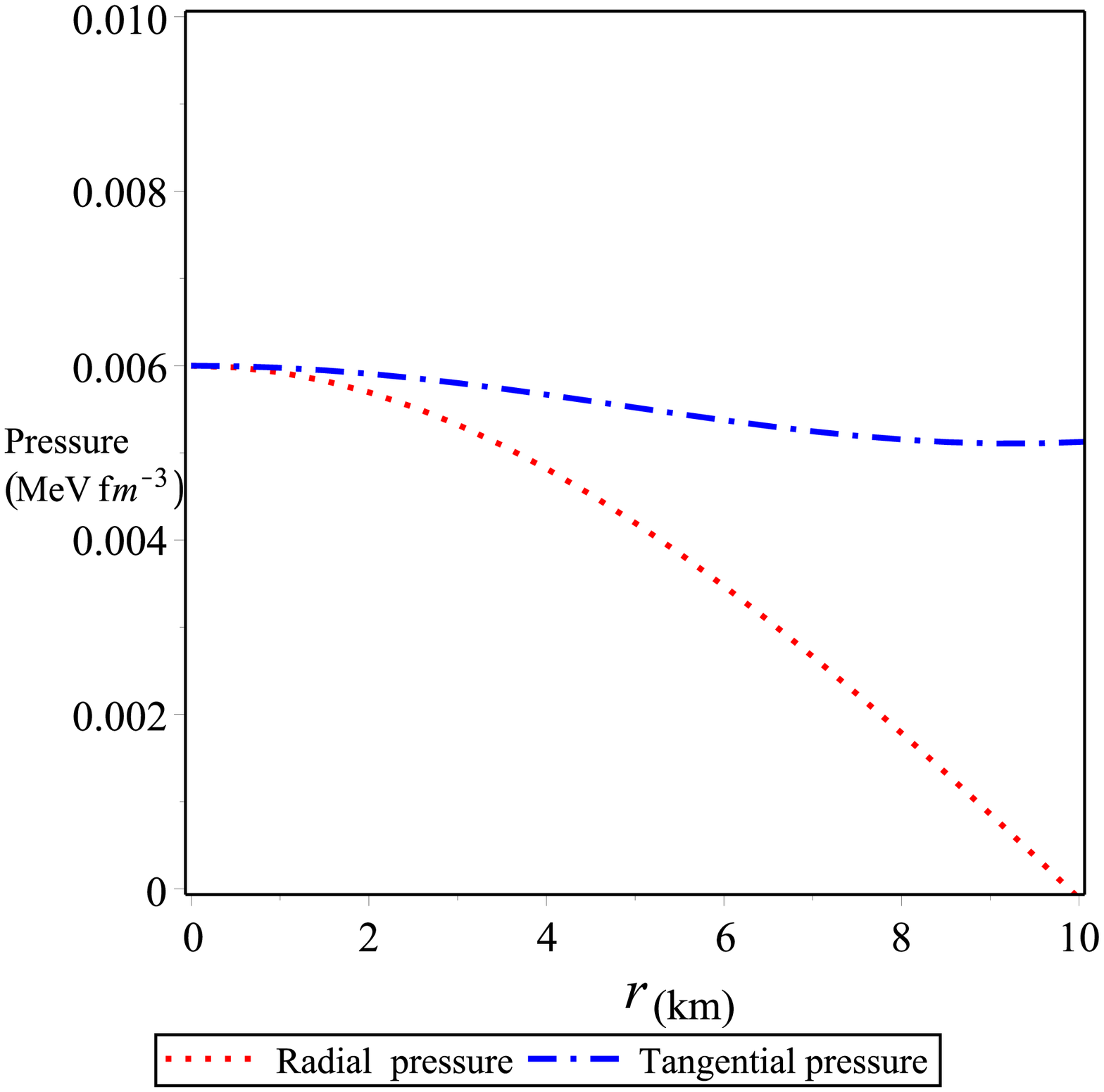}}
\caption[figtopcap]{\small{{Plots of the density, radial and tangential pressures of (\ref{sol})  vs the radial coordinate  $r$ in Km  adopting PSR J 1614--2230 constraints. According to this, we assume the values of  constants as  $k \sim 28.248$, $c_2\sim 0.003$, $k_1\sim 0.492$.}}}
\label{Fig:2}
\end{figure}
Fig.  \ref{Fig:2} shows that density, radial and tangential pressures are positive as required for realistic  stellar configuration. Moreover, as Fig. \ref{Fig:2} \subref{fig:density} shows,    the density is high at the center and  decreases  far from it.   Fig. \ref{Fig:2} \subref{fig:pressure}  shows that the radial pressure  goes to the zero at the boundary while
the tangential pressure remains non-zero at the boundary. Also this feature is relevant for a realistic model.
\begin{figure}
\centering
\subfigure{\label{fig:An}\includegraphics[scale=0.4]{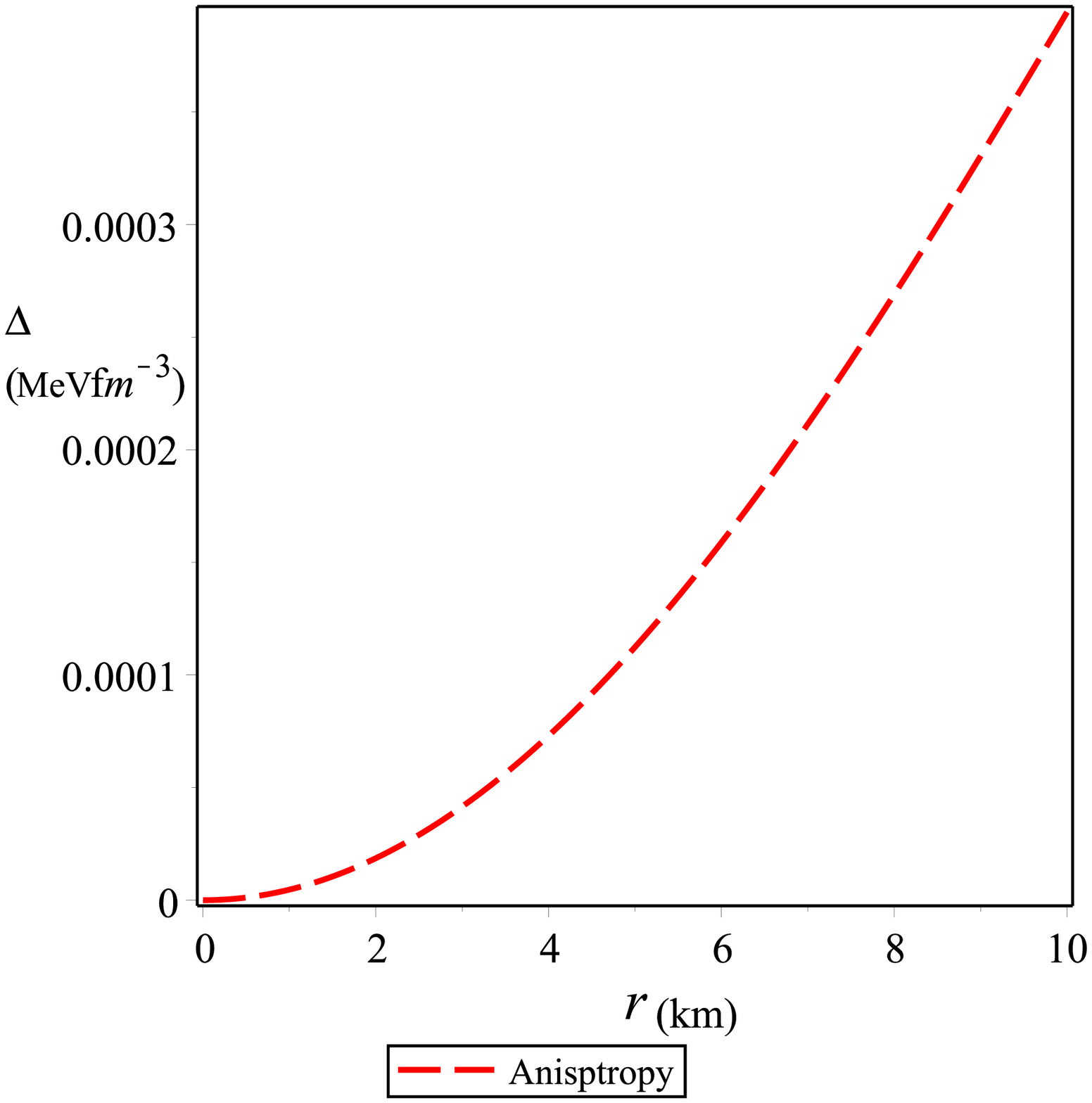}}
\subfigure{\label{fig:An1}\includegraphics[scale=.4]{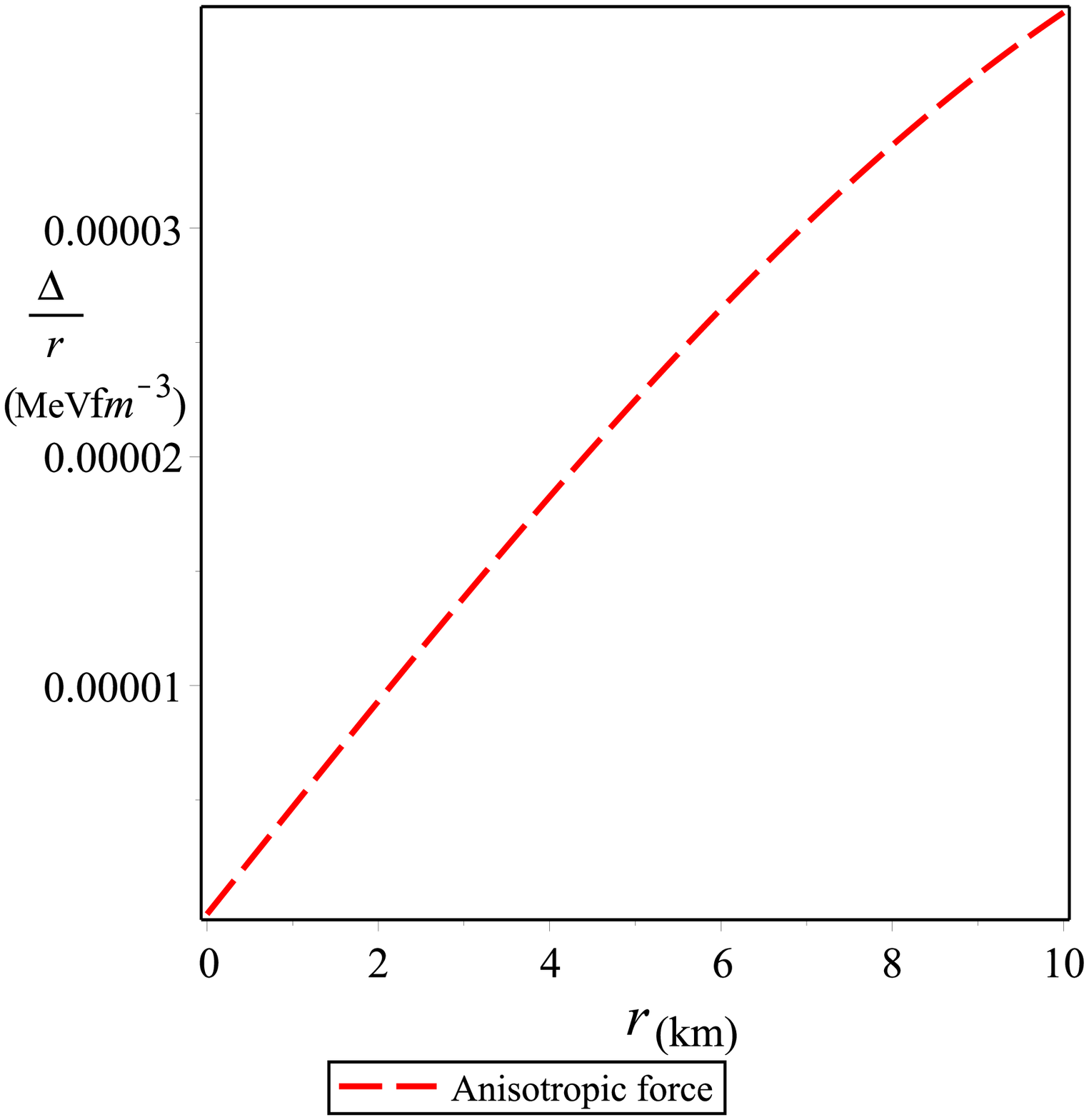}}
\caption[figtopcap]{\small{{{Plots of the  anisotropy and anisotropic forces for solution (\ref{sol})  vs the radial coordinate  $r$ in Km   using  PSR J 1614--2230 constraints.}}}}
\label{Fig:3}
\end{figure}

Fig.  \ref{Fig:3} \subref{fig:An} shows that the anisotropy is vanishing at the center and increases at the surface of the star. Specifically,  Fig.  \ref{Fig:3} \subref{fig:An1} shows that the anisotropic force $\frac{\Delta}{r}$ is positive.  This means that it is repulsive due to the fact that $p_t-p_r>0$.

\begin{figure}
\centering
\includegraphics[scale=0.4]{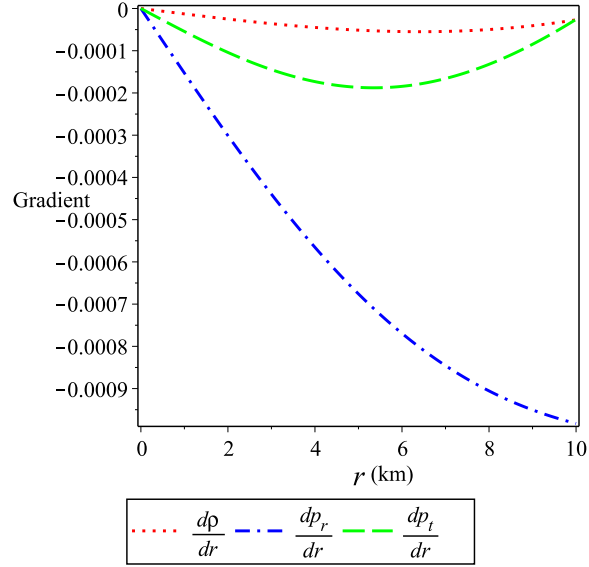}
\caption[figtopcap]{\small{{ Plots of the  density gradient, radial and tangential pressures of solution (\ref{sol}) versus the radial coordinate $r$ in Km    using the  PSR J 1614--2230 constraints.}}}
\label{Fig:4}
\end{figure}

Fig. \ref{Fig:4} shows that the gradients of  density, radial and tangential pressures are negative which confirms the decreasing of density, radial and transverse pressures through the stellar configuration.

\begin{figure}
\centering
\includegraphics[scale=0.4]{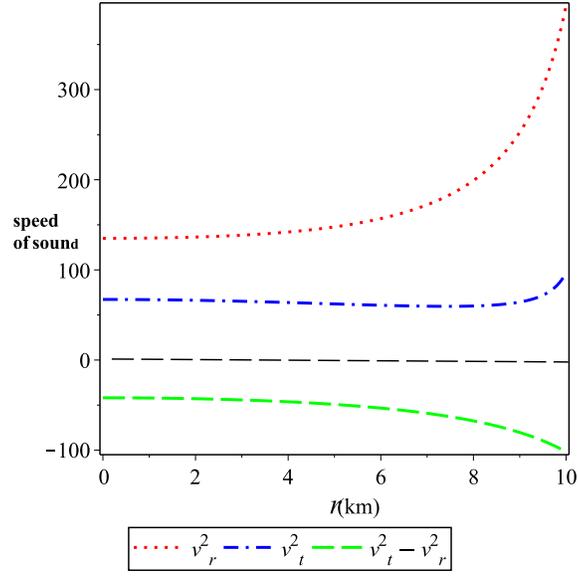}
\caption[figtopcap]{\small{{{Plots of the radial and tangential sound speeds  vs the radial coordinate $r$ in Km using  the PSR J 1614--2230 constraints.}}}}
\label{Fig:5}
\end{figure}
In  Fig. \ref{Fig:5}, the  radial and tangential speeds  of sound are reported. They are  positive and both of them are less than one. This result confirms the non-violations of causality condition in the interior of the star.
\begin{figure}
\centering
\subfigure[~Weak  energy conditions]{\label{fig:WEC}\includegraphics[scale=0.3]{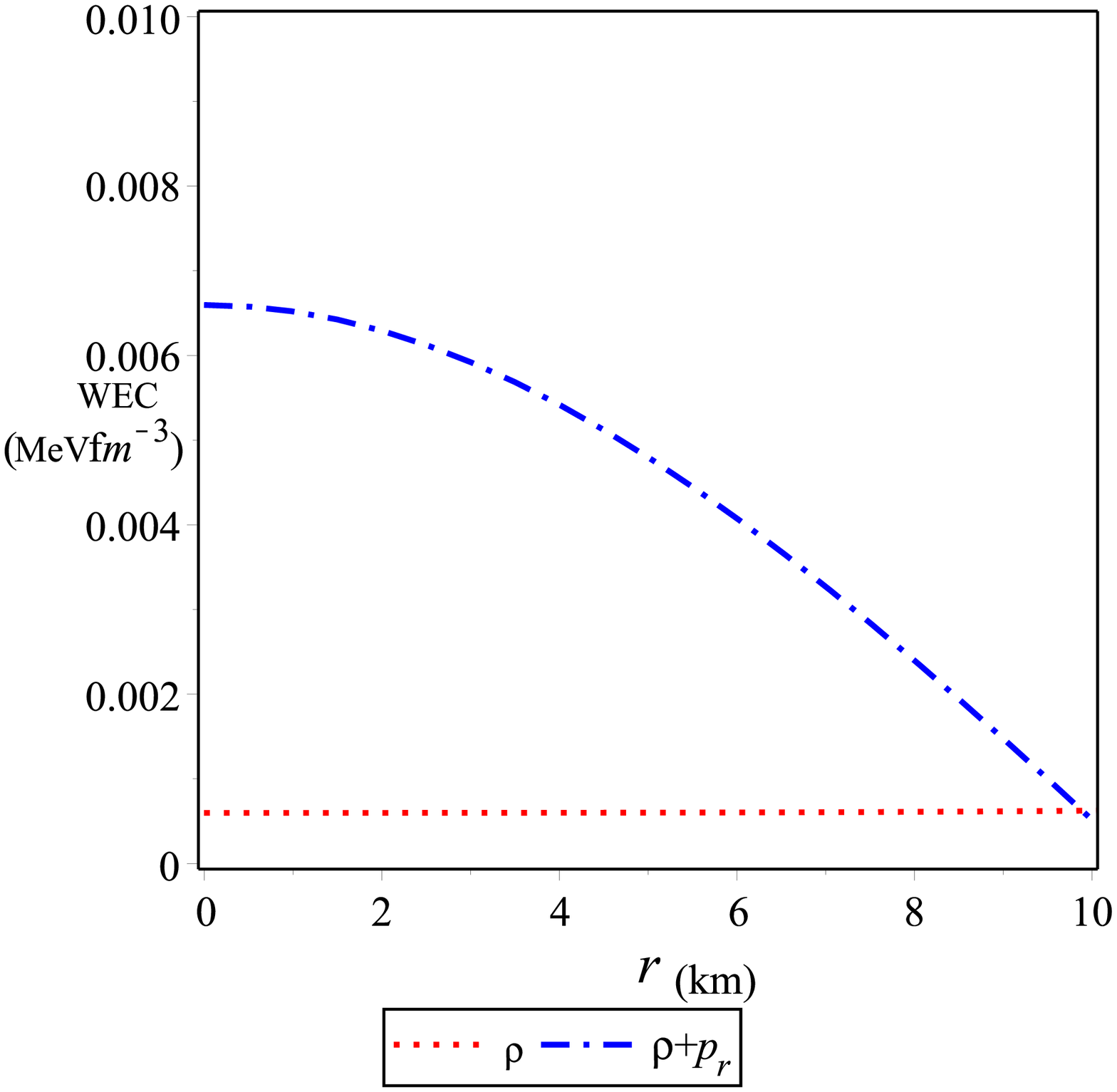}}
\subfigure[~Null  energy conditions]{\label{fig:NEC}\includegraphics[scale=0.3]{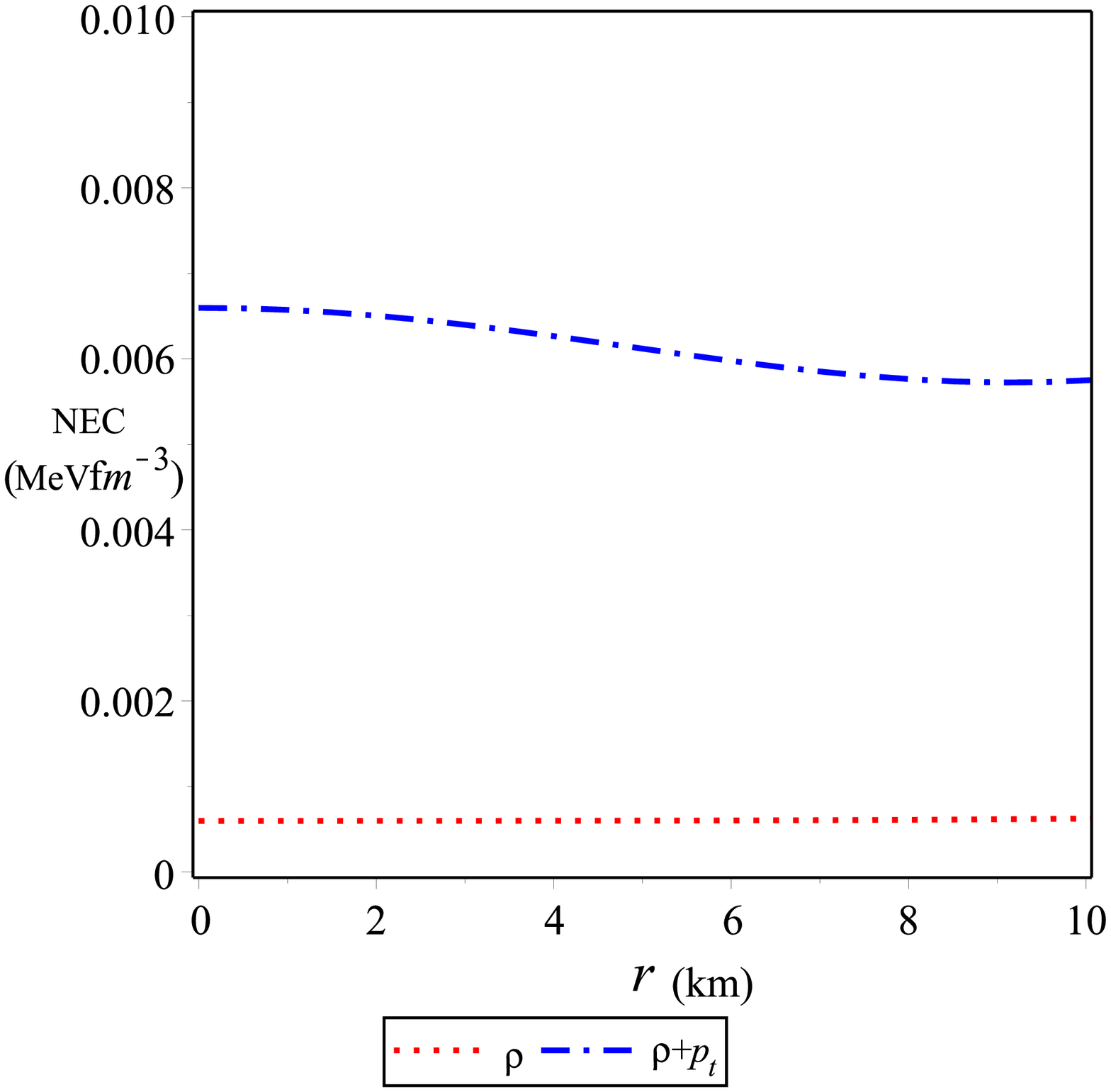}}
\subfigure[~Strong energy condition]{\label{fig:SEC}\includegraphics[scale=.3]{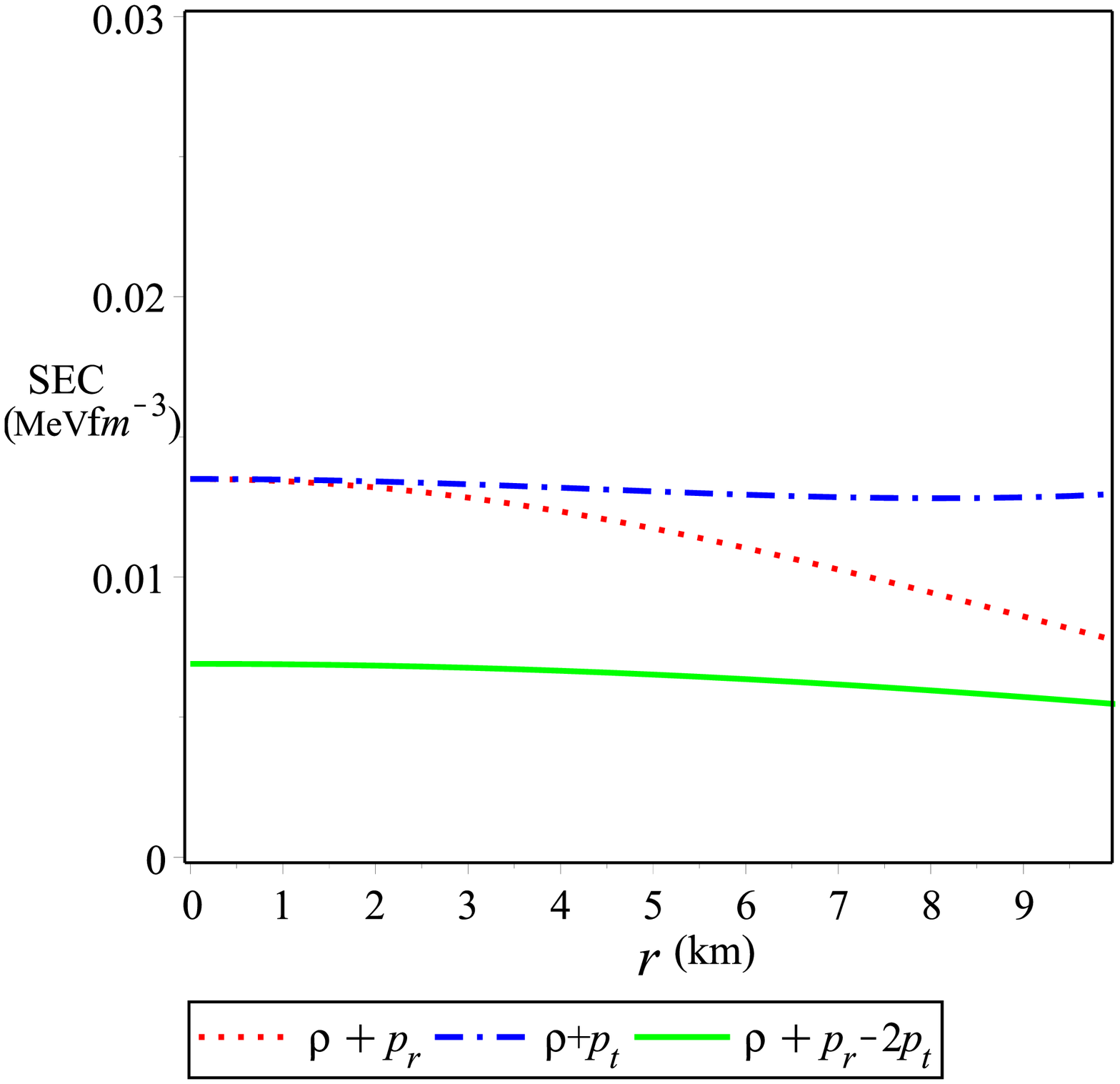}}
\subfigure[~Dominant energy condition]{\label{fig:DEC}\includegraphics[scale=.3]{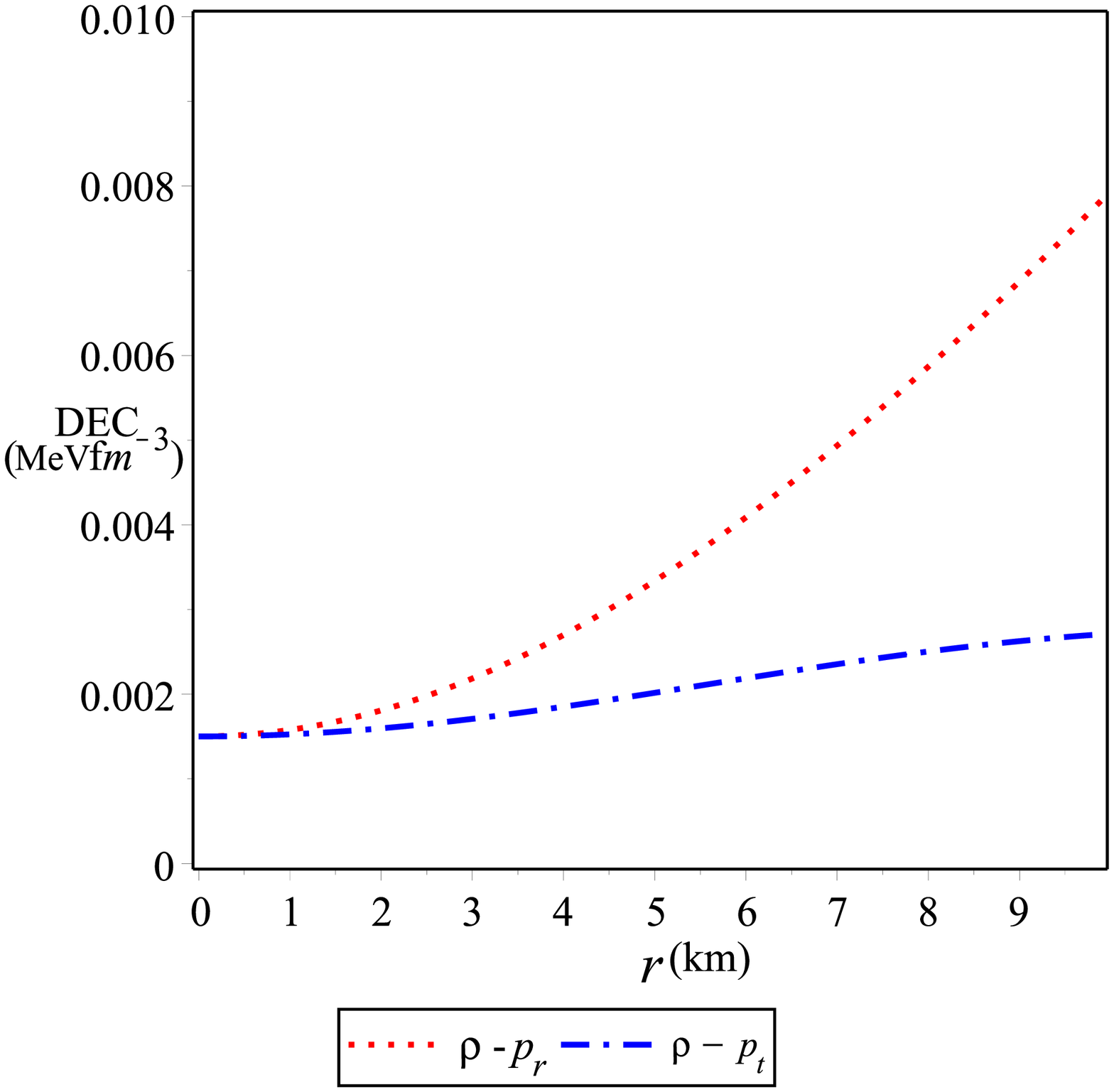}}
\caption[figtopcap]{\small{{{Plots of the  weak, null, strong and dominant energy conditions coming from (\ref{sol}) vs the radial coordinate  $r$ in Km  using  the PSR J 1614--2230 constraints.}}}}
\label{Fig:6}
\end{figure}

Fig. \ref{Fig:6} represent the behavior of the energy conditions. In particular, Figs. \ref{Fig:6} \subref{fig:WEC}, \subref{fig:NEC}, \subref{fig:SEC} and \subref{fig:DEC} show the positive values of the WEC,  NEC,  SEC and DEC energy conditions.   Therefore, all the energy conditions are satisfied throughout the stellar configuration as required for
a physically meaningful stellar model.
\begin{figure}
\centering
\subfigure[~Matching condition of the potential G(R)]{\label{fig:sm}\includegraphics[scale=0.4]{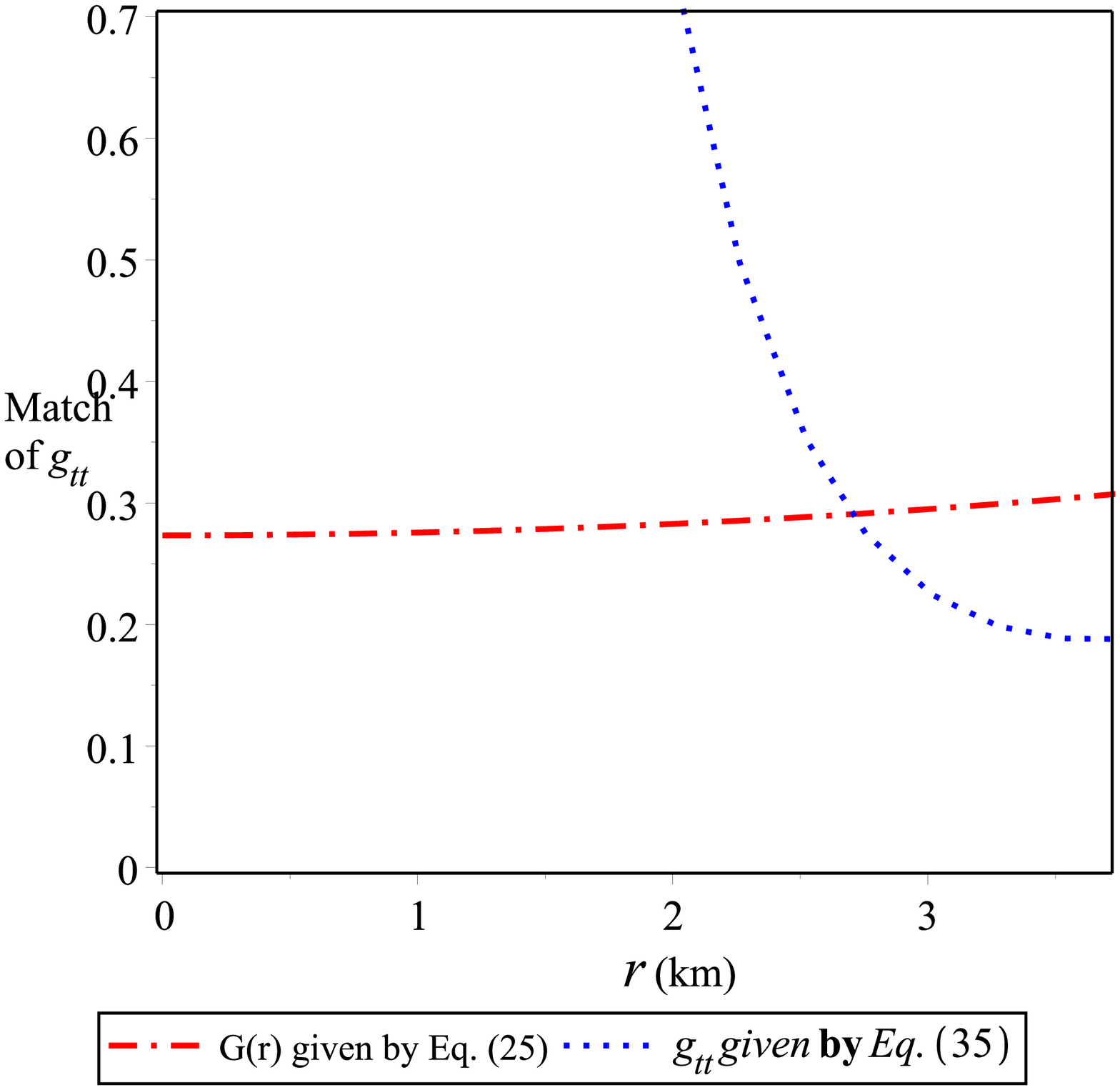}}
\subfigure[~Matching condition of the potential H(R)]{\label{fig:sm1}\includegraphics[scale=.4]{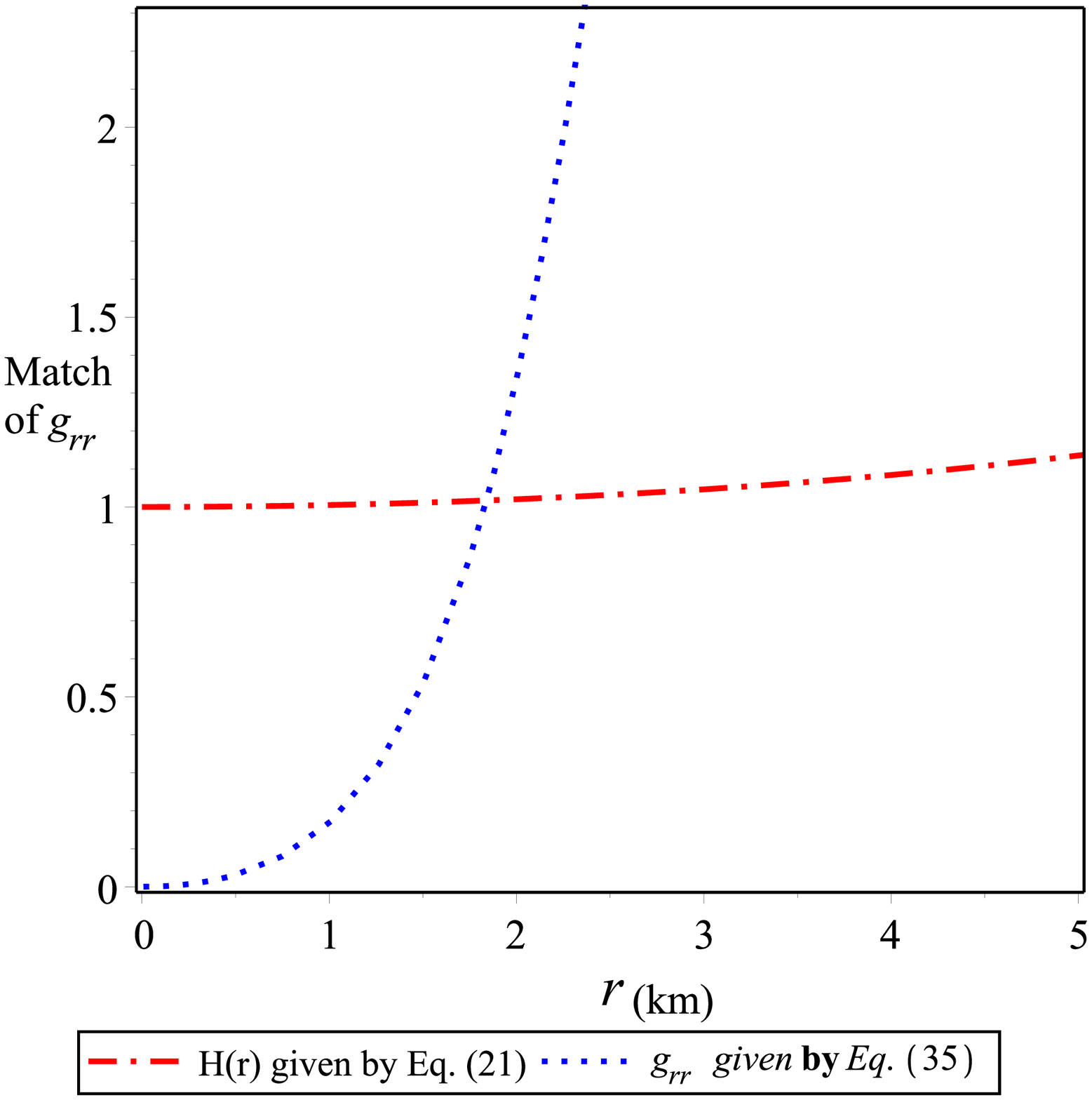}}
\caption[figtopcap]{\small{{Plot of the radial coordinate  $r$ in Km  versus the  weak, null, strong and dominant energy conditions from (\ref{sol})  using the constants constrained by PSR J 1614--2230 data.}}}
\label{Fig:7}
\end{figure}

The matching of the
interior and exterior metrics  at the boundary are shown in Fig. \ref{Fig:7}. Fig. \ref{Fig:7} \subref{fig:sm} represents the smooth matching between $G(R)$ and $1-\frac{2M}{R}+\frac{Q^2}{R^2}$. Fig. \ref{Fig:7} \subref{fig:sm1} shows the smooth matching between $H(R)$ and $(1-\frac{2M}{R}+\frac{Q^2}{R^2})^{-1}$.
\begin{figure}
\centering
\subfigure{\label{fig:pr}\includegraphics[scale=0.4]{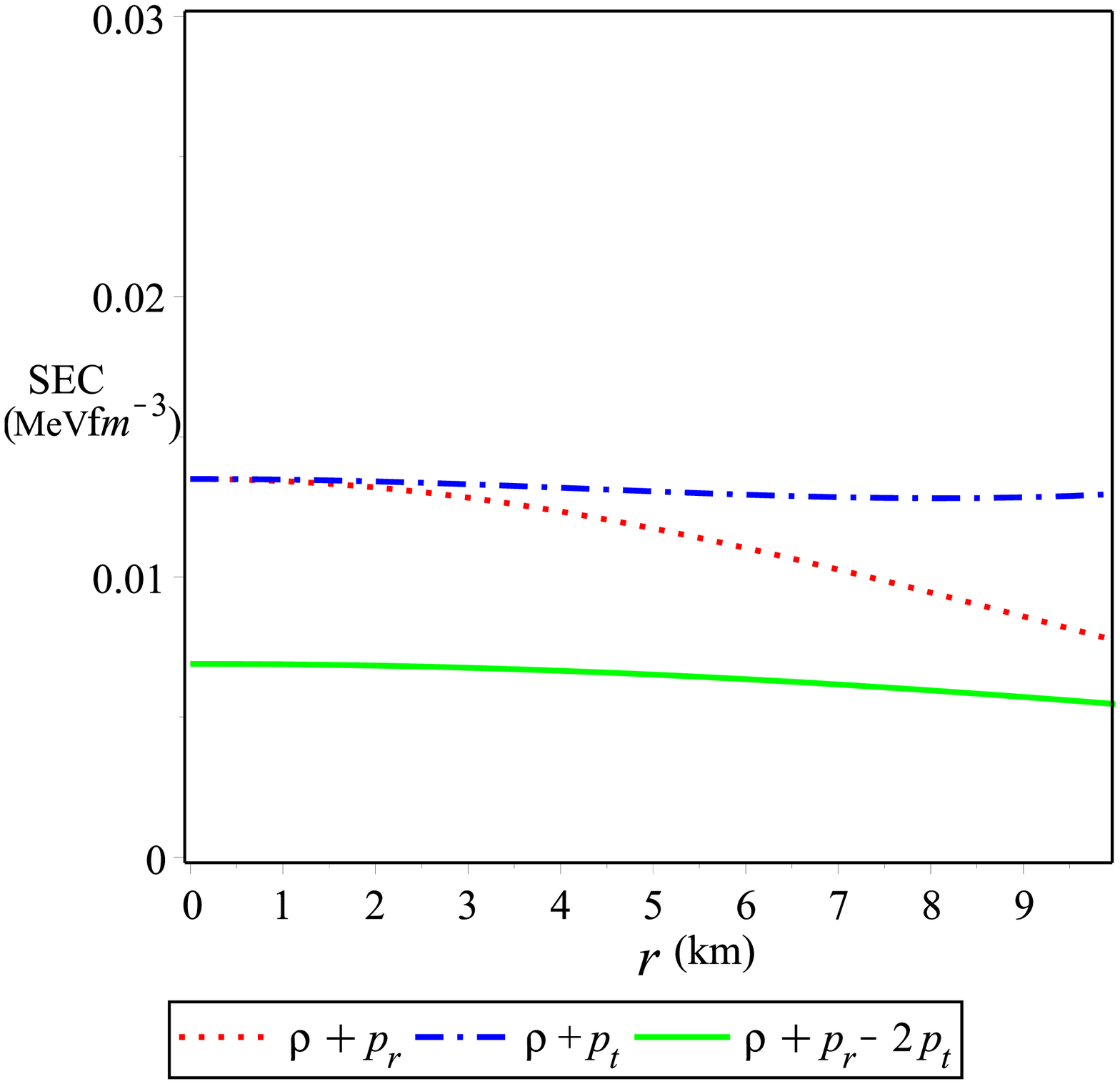}}
\subfigure{\label{fig:pt}\includegraphics[scale=.4]{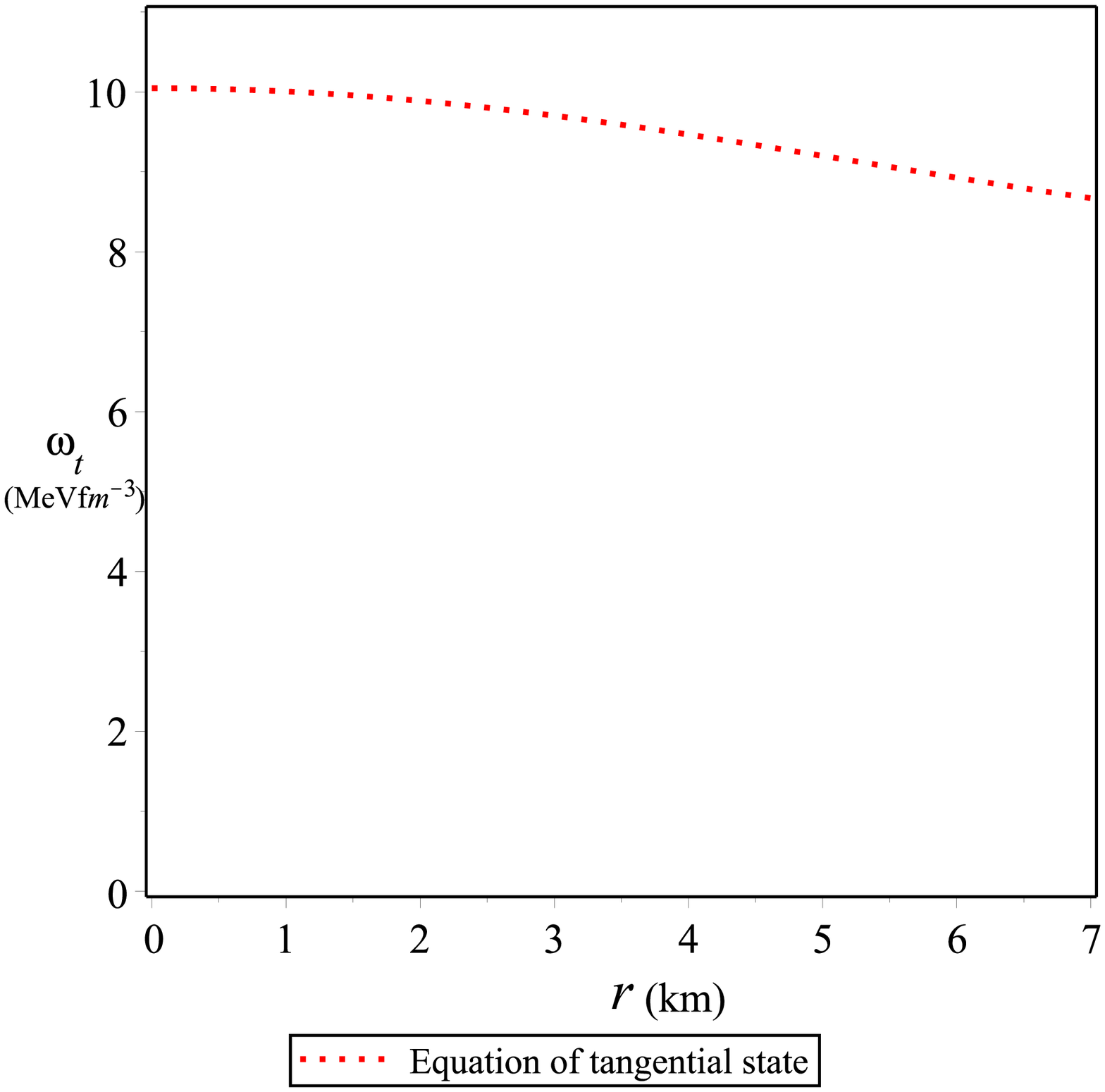}}
\caption[figtopcap]{\small{{Plot of the density $\rho$ in  $MeVfm^3$ versus the  radial pressure $p_r$ and tangential pressure $p_t$ $MeVfm^3$,  using the constants constrained by PSR J 1614--2230.}}}
\label{Fig:8}
\end{figure}

In Fig.  \ref{Fig:8}, we have plotted radial and tangential EoS. As Fig  \ref{Fig:8} \subref{fig:pr}, and \subref{fig:pt} show, the EoS is not  linear. In Ref. \cite{Das:2019dkn}, authors derive the EoS for neutral compact stars and show that it is almost a linear one.  Here, both  the radial and tangential  EoS  show  non-linear form which  is due to the contribution of the electric field.
\begin{figure}
\centering
\includegraphics[scale=0.4]{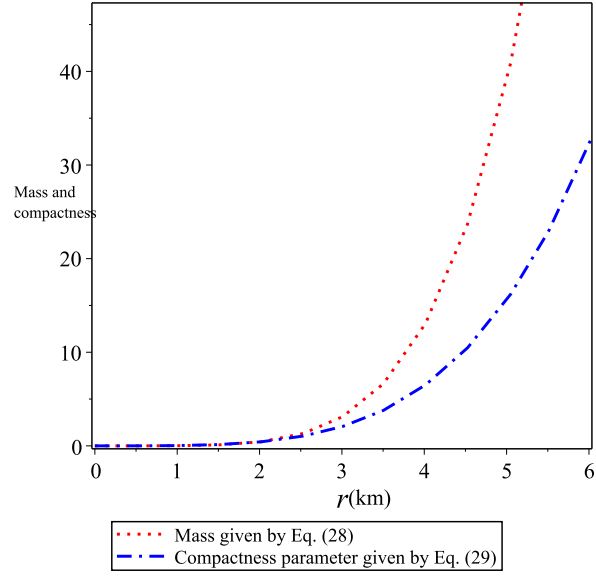}
\caption[figtopcap]{\small{{Plot of the Mass function vs the  radius  $r$  using the constants constrained by PSR J 1614--2230.}}}
\label{Fig:9}
\end{figure}

The mass function  given  by Eq. (\ref{mas}) is plotted in Fig. \ref{Fig:9} which shows that it is a monotonically
increasing function of the radial coordinate and $M(r=0) = 0$.  Moreover, Fig. \ref{Fig:9} shows  the behavior of the compactness parameter of star which is increasing.
\begin{figure}
\centering
\includegraphics[scale=0.4]{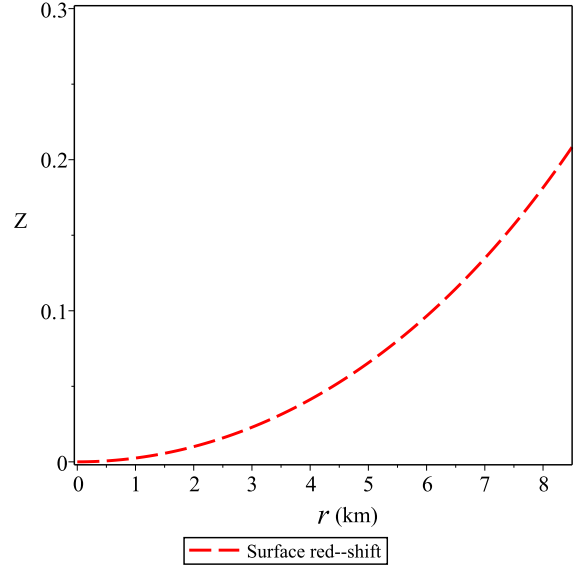}
\caption[figtopcap]{\small{{Plot of the surface red--shift versus the  radius  $r$  using the constants constrained by PSR J 1614--2230.}}}
\label{Fig:10}
\end{figure}

 The radial variation of the surface red-shift is plotted in  Fig.  \ref{Fig:10}.
 B\"{o}hmer and Harko \cite{Bohmer2006} constrained the surface red-shift to be  $Z\leq 5$. The surface redshift of this model is calculated according to $PSR J 1614-2230$ and found to be $0.186927610$.

Using Eqs. (\ref{sol})  and (\ref{mas1}),  it is possible to  derive the Mass-Radius  $(M,R)$ relation of the model  for  a given value of the surface density $(\rho(r = R) =1.5 \approx 10^{15} \frac{g}{ cm^3})$. In   Fig.  \ref{Fig:11}, it is reported  considering  also the compactness--radius relation. As Fig. \ref{Fig:11} \subref{fig:Mass--Radius relation for non-vanishing electromagnetic} shows,  our model has a maximum mass   3$M_\circledcirc$ which is well beyond the recently reported values of 2.50-2.67$M_\circledcirc$ recently reported by the LIGO collaboration \cite{Abbott:2020khf}. This means that anomalous compact objects can be addressed in the framework of TEGR.

\begin{figure}
\centering
\subfigure[~Mass--Radius with vanishing electromagnetic ]{\label{fig:Mass--Radius relation for vanishing electromagnetic}\includegraphics[scale=0.3]{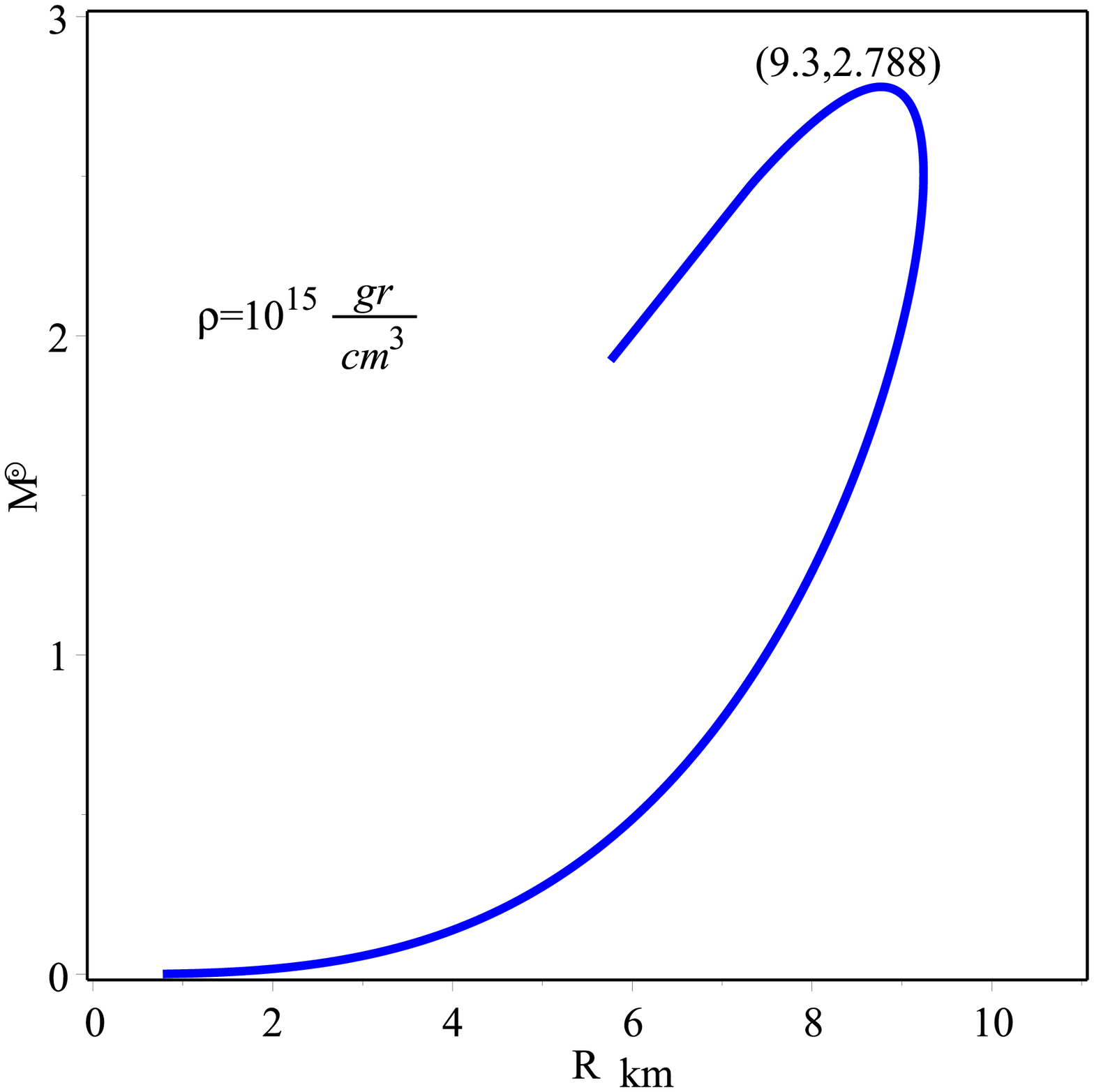}}
\subfigure[~Compactness--Radius with vanishing electromagnetic]{\label{fig:Compactness--Radius relation for vanishing electromagnetic}\includegraphics[scale=0.3]{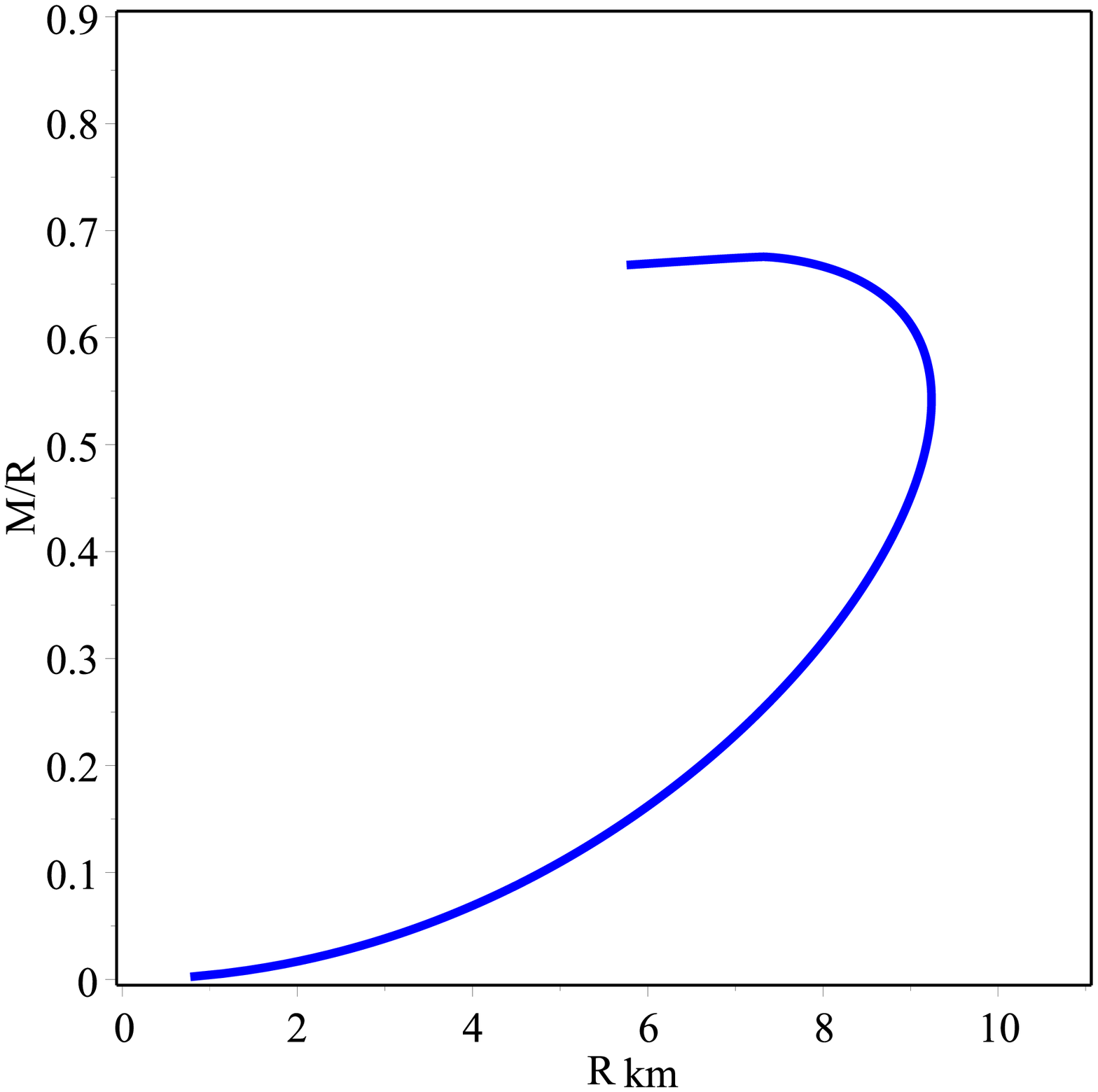}}
\subfigure[~Mass--Radius with electromagnetic ]{\label{fig:Mass--Radius relation for non-vanishing electromagnetic}\includegraphics[scale=.3]{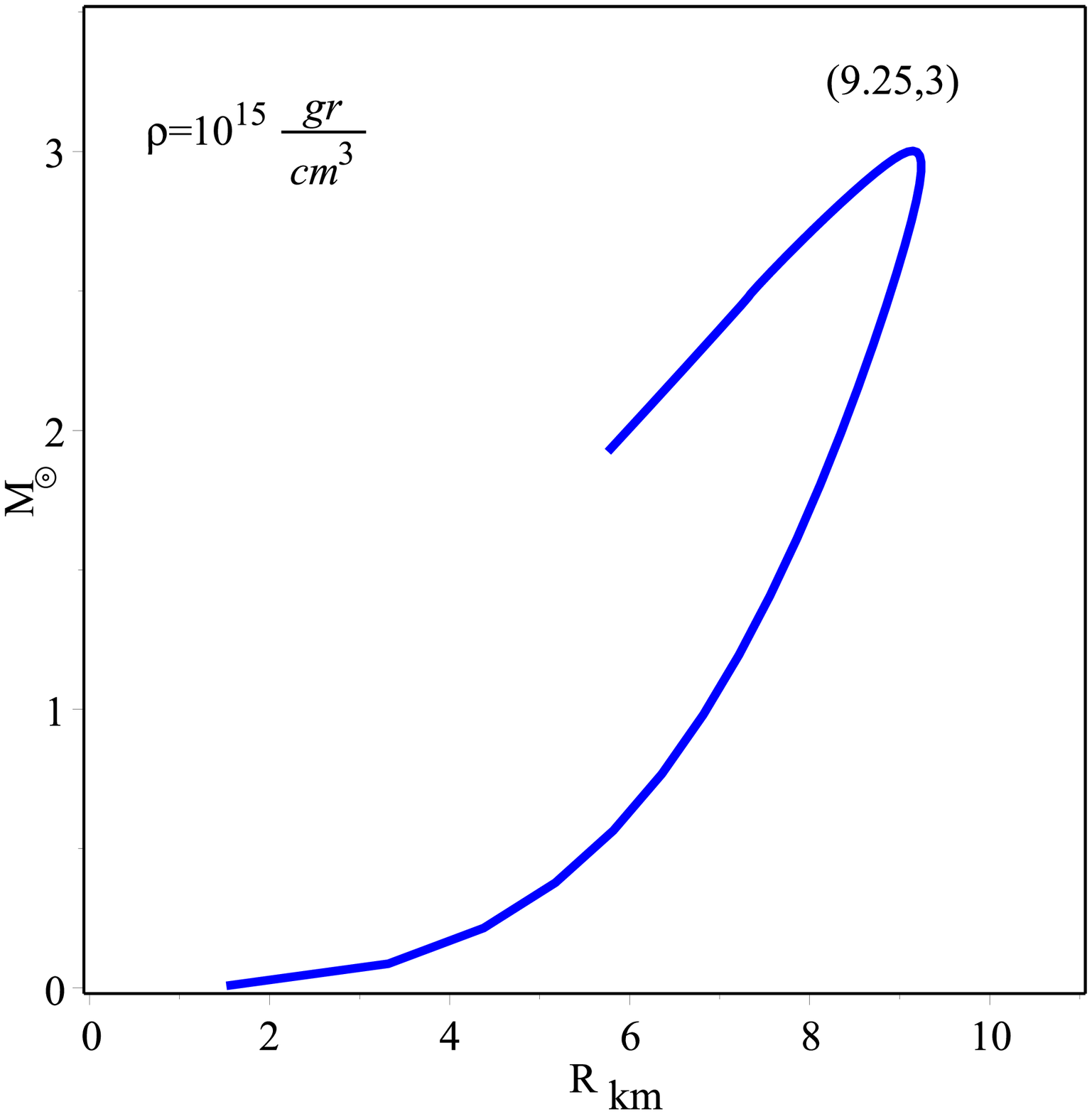}}
\subfigure[~Compactness--Radius with  electromagnetic ]{\label{fig:Compactness--Radius relation for non-vanishing electromagnetic}\includegraphics[scale=.3]{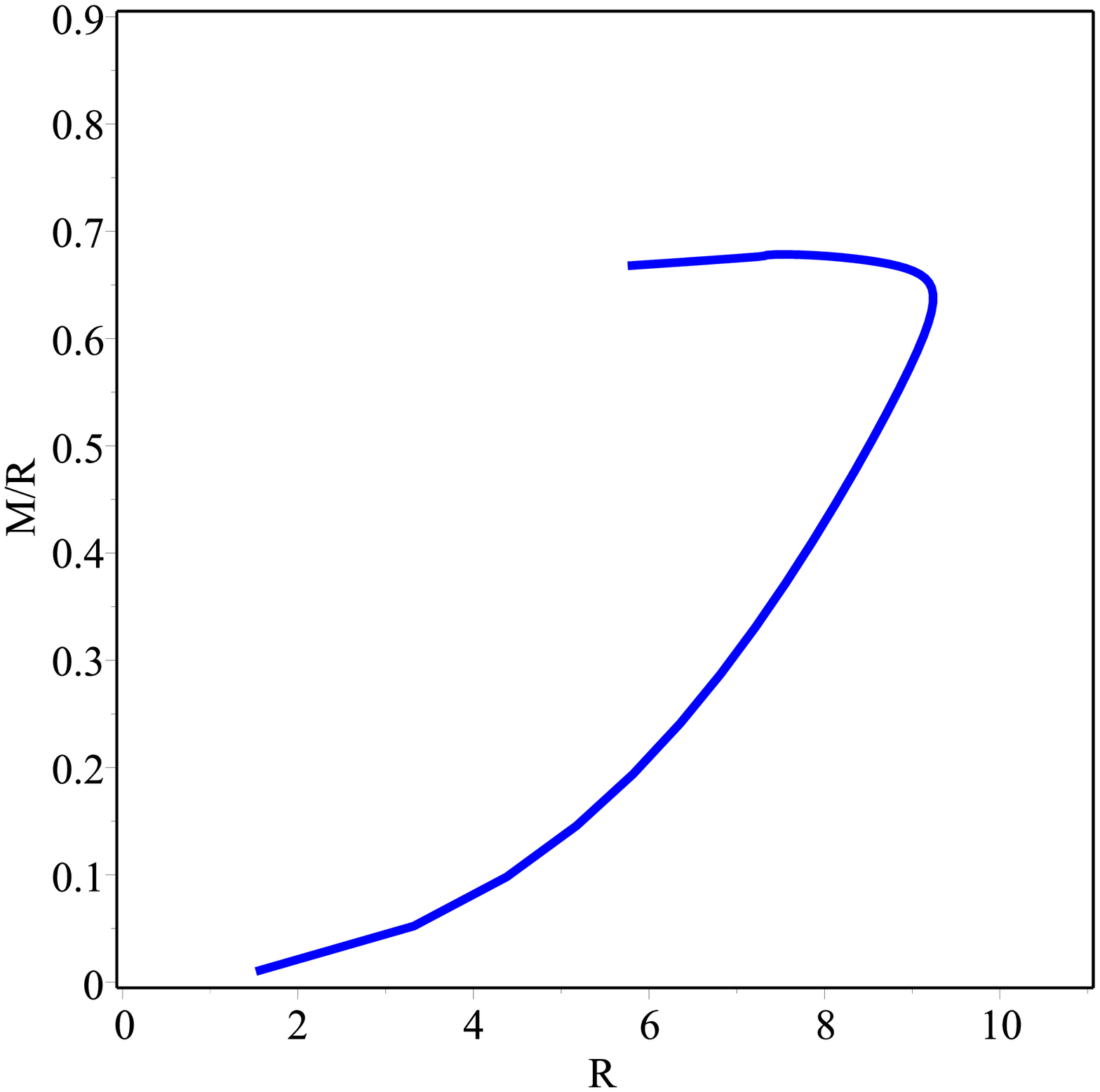}}
\caption[figtopcap]{\small{{Plot of the Mass--Radius and compact--radius with/without electromagnetic field}}}
\label{Fig:11}
\end{figure}
\section{Stability of the model}\label{stability}
In this section we are going to discuss the stability issue using two different techniques; the Tolman-Oppenheimer-Volkoff equations and the adiabatic index.
\subsection{Equilibrium analysis through Tolman-Oppenheimer-Volkoff equation}
In this subsection we are going to discuss the stability of
the   model. To this goal,  we assume hydrostatic equilibrium
through the Tolman-Oppenheimer-Volkoff (TOV) equation.
Using the TOV equation \cite{PhysRev.55.364,PhysRev.55.374} as that presented  in \cite{PoncedeLeon1993}, we get the following  form
\begin{eqnarray}\label{TOV}   \frac{2[p_t-p_r]}{r}-\frac{M_g(r)[\rho(r)+p_r]\sqrt{G}}{r\sqrt{H}}-\frac{dp_r}{r}+\sigma E\sqrt{H}=0,
 \end{eqnarray}
with $M_g(r)$ being  the gravitational mass at
radius $r$, as defined by the  Tolman-Whittaker mass
formula which gives
\begin{eqnarray}\label{ma}   M_g(r)=4\pi{\int_0}^r\Big({T_t}^t-{T_r}^r-{T_\theta}^\theta-{T_\phi}^\phi\Big)r^2\sqrt{GH}dr=\frac{rG'\sqrt{H}}{2G}\,,
 \end{eqnarray}
Inserting Eq. (\ref{ma}) into (\ref{TOV}),  we get
\begin{eqnarray}\label{ma1}  \frac{2(p_t-p_r)}{r}-\frac{dp_r}{dr}-\frac{G'[\rho(r)+p_r]}{2\sqrt{G}}+\sigma E\sqrt{H}=F_g+F_a+F_h+F_e=0\,,
 \end{eqnarray}
 where $F_g=-\frac{G'[\rho(r)+p_r]}{2{G}}$,  $F_a=\frac{2(p_t-p_r)}{r}$, $F_h=-\frac{dp_r}{dr}$ and $F_e=\sigma E\sqrt{H}$  are the gravitational,  the anisotropic,  the hydrostatic and the electromagnetic forces respectively. The behavior of the TOV equation for the  model  (\ref{sol}) is shown in Fig. \ref{Fig:12}.
\begin{figure}
\centering
\includegraphics[scale=0.4]{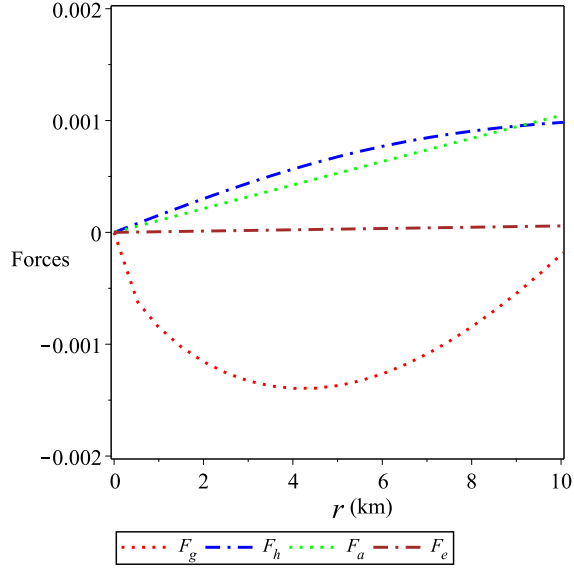}
\caption[figtopcap]{\small{{Plot of the gravitational,  the anisotropic,  the hydrostatic and the electromagnetic forces  versus the  radius  $r$  using the constants constrained from $\textit {PSR J 1614--2230}$.}}}
\label{Fig:12}
\end{figure}
The four different forces are plotted in Fig. \ref{Fig:12}.  It
shows that hydrostatics, anisotropic and electromagnetic forces are positive and
 dominated by the gravitational force which is negative to
keep the system in static equilibrium.

\subsection{Stability in the static state}
It was shown by Harrison, Zeldovich and Novikov \cite{1971reas.book.....Z, Zeldovich:1983cr} that, for stable compact stars, 
the gradient of  central density, with  regard to the mass increasing,  must be positive, i.e., $\frac{\partial M}{\partial \rho_{_{r_0}}}> 0$. If this condition is verified,  we have stable configurations. Specifically,  stable or unstable regions  are separated 
when we have a constant mass i.e. $\frac{\partial M}{\partial \rho_{r_0}}= 0$. Now we are going to apply this condition to our solution (\ref{sol}). For this purpose,  we calculate the  central density for  solution (\ref{sol})  and get
\begin{eqnarray} \label{sta} \rho_{_{r_0}}=\frac{3}{2\pi k^2} \Rightarrow k=\pm\frac{\sqrt{3}}{ \pi \sqrt{2\rho_{_{r_0}}}}\,.\end{eqnarray}
Using Eq.~(\ref{sta}) in Eq.~(\ref{mas1}),  we get
 \begin{eqnarray}\label{sta1}
 M(\rho_{_{r_0}})=\frac{R^3(81R^2k_1{}^2+\pi[324 k_1R^2+540 \rho_{_{r_0}}-540\pi R^2\rho_{_{r_0}}{}^2+240\pi^2R^4\rho_{_{r_0}}{}^3-40\pi^3R^4\rho_{_{r_0}}{}^4])}{405} \,.
 \end{eqnarray}
With the help of Eq. (\ref{sta1}),  we have
 \begin{eqnarray} \label{sta2} \frac{\partial M}{\partial \rho_{_{r_0}}}=\frac{\pi R^3(540 -1080\pi R^2\rho_{_{r_0}}+720\pi^2R^4\rho_{_{r_0}}{}^2-160\pi^3R^6\rho_{_{r_0}}{}^3)}{405} \,.
 \end{eqnarray}
From Eq. (\ref{sta2}), it is crear that  solution (\ref{sol}) has a stable configuration since  $\frac{\partial M}{\partial \rho_{r_0}}> 0$. The behavior of (\ref{sta1}) and (\ref{sta2}) are show in Fig. \ref{Fig:13}. Fig. \ref{Fig:13} shows that  mass increases as the energy density increases and the mass gradient  decreases as energy density increases.
\begin{figure}[!ht]
\centering
{\label{fig:st}\includegraphics[scale=.45]{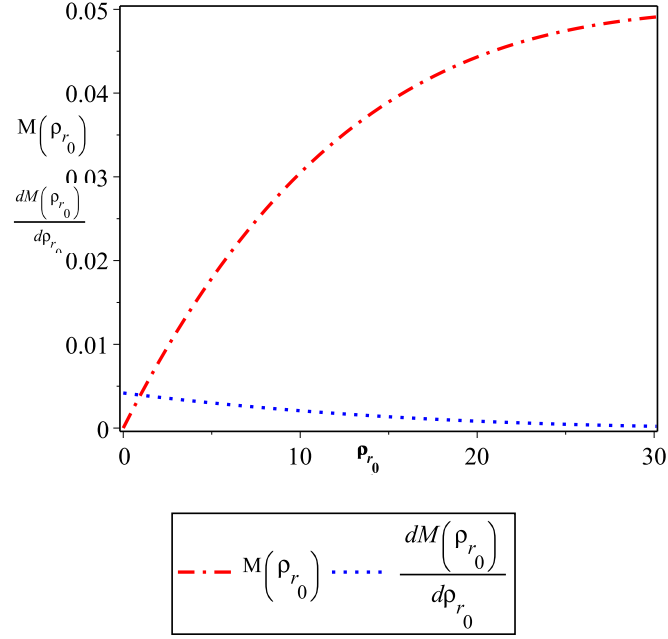}}
\caption[figtopcap]{\small{{Static stability of  (\ref{sol}) against  $\rho_{r_0}$ in $\mathrm{km}^{-3}$ when $b_0=b_2=0.0001, R=0.1$.}}}
\label{Fig:13}
\end{figure}
\subsection{Adiabatic index}
The stable equilibrium
configuration of a spherically symmetric system can be studied  using the adiabatic index which is a basic ingredient of
the stability  criterion. Let us  consider an adiabatic
perturbation, the adiabatic index $\Gamma$, is defined as \cite{1964ApJ...140..417C,1989A&A...221....4M,10.1093/mnras/265.3.533}
\begin{eqnarray}\label{a11}  \Gamma=\left(\frac{\rho+p}{p}\right)\left(\frac{dp}{d\rho}\right)\,.
 \end{eqnarray}
 A Newtonian isotropic sphere is in stable
equilibrium if the adiabatic index $\Gamma>\frac{4}{3}$
as reported in  Heintzmann
and Hillebrandth \cite{1975A&A....38...51H}. For $\Gamma=\frac{4}{3}$, the  isotropic
sphere is  in neutral equilibrium. Based on some  works by Chan et al. \cite{10.1093/mnras/265.3.533},  one can require  the following
condition for the stability of a relativistic anisotropic sphere $\Gamma >\gamma$ where \begin{eqnarray}\label{ai}  \gamma=\frac{4}{3}-\left\{\frac{4(p_r-p_t)}{3\lvert p'_r\lvert}\right\}_{max}\,.
 \end{eqnarray}
 Using Eq. (\ref{ai}),  we get
 \begin{eqnarray}\label{a12}  &&\Gamma=\frac{4}{3}-\frac{2}{3}\Big[(c_2r^2+1)^2(4k_1{}^2k^8+6k^4-8r^2k^2+3r^4)\Big]\times\Big[2(c_2r^2+1)^2k^8k_1{}^2+16k^6c_2-18c_2{}^2r^4k^4-36k^4c_2r^2+6k^4\nonumber\\
 &&+16c_2r^4k^2-16k^2r^2+16c_2{}^2r^6k^2+2c_2r^6+9r^4-3c_2{}^2r^8+4c_2{}^2k^8\Big]^{-1}.
 \end{eqnarray}
 \begin{figure}
\centering
\includegraphics[scale=0.4]{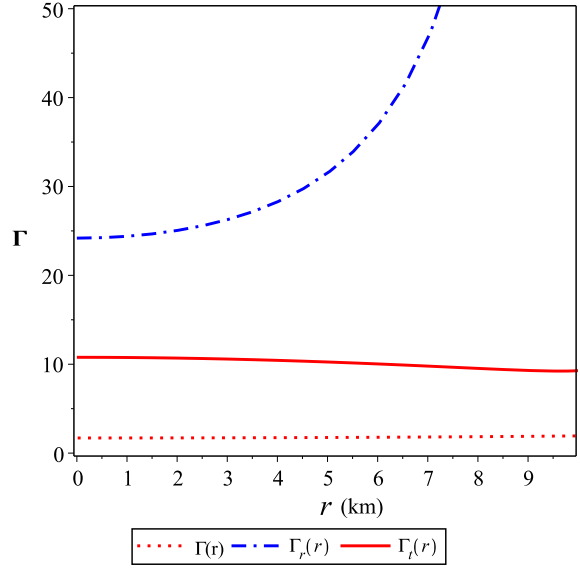}
\caption[figtopcap]{\small{{Plot of the adiabatic index versus the  radius  $r$  using the constants constrained from PSR J 1614--2230.}}}
\label{Fig:14}
\end{figure}
 From Eq. (\ref{a11}),  we obtain  the adiabatic index of solution (\ref{sol})   in the form
 \begin{eqnarray}\label{aic} &&\Gamma_r=\Big\{4(k^2-r^2)^3(3+c_2[2r^2+k^2])\Big[2k^8(k_1{}^2\{1+c_2{}r^2\}^2+2c_2{}^2)+16c_2k^6+6k^4(1-3c_2r^2[2+c_2r^2])\nonumber\\
 &&+16k^2r^2(c_2r^2-1+c_2{}^2r^4)-3c_2{}^2r^8
 +2c_2r^6+9r^4\Big]\Big\}\Big\{[1+c_2r^2]^2\Big[2k^8(k_1{}^2r^2\{1+c_2r^2\}-2c_2{}^2)+16c_2k^6r^2\nonumber\\
 &&+6k^4r^2[1-3c_2r^2]+8k^2r^2(c_2r^2-1)+3r^6
-c_2r^8\Big][2k^8k_1{}^2-30k^4+56k^2r^2-27r^4]\Big\}\,, \nonumber\\
 &&\Gamma_t=\Big[4k^8(k_1{}^2r^2\{1+c_2{}r^2\}+c_2)+4k^6(3-c_2r^2)-6k^4r^2(5+c_2r^2)+4k^2r^2(3c_2r^4+7)-9r^6-5c_2r^8\Big]\nonumber\\
 &&\Big[k^8(k_1{}^2\{1+c_2r^2\}^2-2c_2{}^2)-8c_2k^6+12c_2r^2k^4[c_2r^2+2]-8k^2r^4c_2[2c_2r^2+3]+2c_2r^6(4+3r^2c_2)\Big]
 \nonumber\\
 &&\Big\{[1+c_2r^2]^2\Big[k^8(k_1{}^2r^2\{1+c_2r^2\}+2c_2{})-8c_2k^6r^2+12k^4r^4c_2-8k^2r^6c_2+2c_2r^8\Big][2k^8k_1{}^2-30k^4+56k^2r^2-27r^4]\Big\}\,. \nonumber\\
 &&
 \end{eqnarray}

In Fig. \ref{Fig:14} we have reported  $\Gamma$, $\Gamma_r$ and  $\Gamma_t$   respectively.
As it is clear from these plots, it can be seen that the values of $\Gamma_r$ and  $\Gamma_t$  are greater
than $\Gamma$ throughout the stellar interior and hence the stability
condition is fulfilled.
\begin{table*}[t!]
\caption{\label{Table1}%
Values of model parameters}
\begin{ruledtabular}
\begin{tabular*}{\textwidth}{lccccccc}
{{Pulsar}}                              & Mass ($M_{\odot}$) &      {Radius (km)} &   {k}  &    {$k_1$}    & {$c_1$} & {$c_2$}&\\ \hline
&&&&&&&\\
Her X-1                                 &  $0.85\pm 0.15$    &      $8.1\pm0.41$ &   33.43508140    &    0.8680552960$\times10^{-2}$         & 450680.0914 &0.5000000015$\times10^{-2}$     \\
Cen X-3                                &  $1.49\pm 0.08$    &      $9.178\pm0.13$ &   29.76976530    &     0.7072954467$\times 10^{-2}$        &  206264.0176  &0.5000000001$\times10^{-2}$      \\
RX J 1856 -37                           &  $0.9\pm 0.2$      &      $\simeq 6$   &  18.26492898     &    .1121853911$\times10^{-1}$        & 40286.21339      &0.5000000003$\times10^{-2}$  \\
4U1608 - 52                        &  $1.74\pm 0.14$      &      $9.52\pm0.15$   &   28.41404292     &    0.6372911235$\times10^{-2}$         & 144607.5846        &0.4999999996$\times10^{-2}$  \\
EXO 1785 - 248                        &  $1.3\pm 0.2$      &      $8.849\pm0.4$   &  28.50683773     &   0.6625295030$\times10^{-2}$        & 269103.9598         &0.3000000010$\times10^{-2}$ \\
4U1820 - 30                        &  $1.58$      &      $9.1$   &  28.61239325     &    0.2556454675$\times10^{-2}$         & 315783.9636        &0.7000000014$\times10^{-3}$ \\
&&&&&&&\\

\end{tabular*}
\end{ruledtabular}
\end{table*}
\begin{table*}[t!]
\caption{\label{Table1}%
Values of physical quantities}
\begin{ruledtabular}
\begin{tabular*}{\textwidth}{lcccccccccc}
{{Pulsar}}                              &{$\rho\lvert_{_{_{0}}}$} &      {$\rho\lvert_{_{_{R}}}$} &   {$\frac{dp_r}{d\rho}\lvert_{_{_{0}}}$}  &    {$\frac{dp_r}{d\rho}\lvert_{_{_{R}}}$}
  & {$\frac{dp_t}{d\rho}\lvert_{_{_{0}}}$} & {$\frac{dp_t}{d\rho}\lvert_{_{_{R}}}$}&{$(\rho-p_r-2p_t)\lvert_{_{_{0}}}$}&{$(\rho-p_r-2p_t)\lvert_{_{_{R}}}$}&{$z\lvert_{_{_{R}}}$}&\\ \hline
&&&&&&&&&&\\
Her X-1                        &0.54$\times10^{-2}$     & .46$\times10^{-2}$  &   .64   &   .60      & .30&.25&.24$\times10^{-2}$&.25$\times10^{-2}$& .31  \\
Cen X-3                                & 0.68$\times10^{-2}$     &.53$\times10^{-2}$  &  .50  &   .45        &.24&.2 &.38$\times10^{-2}$&.37$\times10^{-2}$& .51   \\
RX J 1856 -37                   &.18$\times10^{-1}$&.14$\times10^{-1}$&.31  &.28 &.84$\times10^{-1}$&.69$\times10^{-1}$ &.15$\times10^{-1}$&.13$\times10^{-1}$&.58    \\
4U1608 - 52                     & .74$\times10^{-2}$   &  .55$\times10^{-2}$     & .44  & .37&.24 &.2&.44$\times10^{-2}$&.41$\times10^{-2}$&.64    \\
EXO 1785 - 248                     & .74$\times10^{-2}$     & .6$\times10^{-2}$    &.36      &.34  & .12&.10 &.56$\times10^{-2}$&.5$\times10^{-2}$& .44 \\
4U1820 - 30                             &.73$\times10^{-2}$ &  .57$\times10^{-2}$  & .24 & .21 &.28$\times10^{-1}$&.24$\times10^{-1}$ &.7$\times10^{-2}$&.7$\times10^{-2}$&.5    \\

\end{tabular*}
\end{ruledtabular}
\end{table*}

Besides PSR J 1614-2230, a similar analysis can be developed for other pulsars. In Table I and II, we report the results for other observed systems.

\section{Discussion and conclusions}\label{S5}
An important remark is in order at this point. It is well known that  TEGR theory is equivalent  to GR up to a total derivative term \cite{Golovnev_2017,Kr_k_2019,Bejarano_2019}.  In this paper, we  considered a combination of  anisotropy and charge in TEGR equations. This situation  gives rise to the effects of enhancing mass and modifying the $(M,R)$ relation of GR.

 For a class of metric potentials and anisotropy functions, we derived an  exact solution capable of  figuring out realistic   compact star configurations. The regularity conditions of the solution at the origin as well as at the surface of the star  show a well  behavior throughout the stellar structure. This is different from  the results reported in  \cite{Singh:2019ykp} for GR. In that study, the authors show that pressure increases outward which is a non-physical situation. The difference between our results and those reported   in  \cite{Singh:2019ykp} is due to  the anisotropy  given in Eq. (\ref{d2}), the presence of charge, and the non-vanishing of the radial pressure.  In  \cite{Singh:2019ykp},   they assumed a vanishing  radial pressure. In our case, we showed that density, radial and tangential pressures behave regularly according to the observational data of the pulsar $\textit {PSR J 1614--2230}$, the first reported  very  massive  neutron star, whose existence   ruled out many EoS \cite{Demorest}.  In order to explain such a system,  exotic matter such as hyperons and kaon condensates, alternative theories of gravity and other hypotheses have been invoked (see e.g.\cite{Astashenok1,Astashenok2,Astashenok3}). The approach seems to work for other systems, as  reported in Tables I and II.

Furthermore,  we show that the anisotropy has a positive value which can be interpreted as  a repulsive force.  This fact  is because  the tangential pressure is greater than the radial pressure, i.e., $p_t>p_r $ \cite{2019JApA...40....8S}.  The issue of stability is studied and we showed that the derived model is stable against the different forces (gravitational, hydrostatic, anisotropic and electromagnetic) acting on it. We also calculated the sound of speed and showed that it is consistent with realistic compact stars in contrast to the analogue charged models formulated   in  GR, where   an  imaginary sound speed is derived \cite{Singh:2019ykp}. Finally we calculated the  adiabatic index  of our model and showed that also it represents a realistic physical  star. It is worth noticing that  the adiabatic index presented in \cite{Singh:2019ykp} has a negative value which is not consistent with  realistic stellar models. This indicates, in a clear way, that our assumption of the metric potential (\ref{pot}) and the anisotropy form (\ref{d2}) are physical assumptions that makes the resulting stellar model consistent with real stellar objects.

We tested the model over a  wide range of reported
observed values of masses and radii of pulsars (Tables I and II). The conclusion is that the  fit is good also in these cases. Finally, we drew the mass--radius relation and showed  the effect of  electric field on it.

It is  shown that the electric charge plays a central role in improving the results  compared with the neutral   case. Among these improvements, we have: \\ i-  In the neutral case, one gets a maximum  mass as $2.788M_{\odot}$ while, in the charged case, the maximum mass becomes $3M_{\odot}$.\\ii- In the neutral case,  one  get an increasing radial pressure as reported in \cite{Singh:2019ykp} while in the charged case, one gets a decreasing one which describes a consistent compact star. \\iii- In the neutral star, one  gets an imaginary  sound speed \cite{Singh:2019ykp} while, in the charged case, we get a real physical  sound speed as shown in Fig. 5.

The approach can be summarized as follows: we  used a non-diagonal form of the tetrad field that  gives  a null value of the torsion as soon as the metric potentials approach to $1$. This  is a necessary condition for any physical tetrad field  as reported in various studies \cite{PhysRevD.98.064047,Abbas:2015yma,Momeni:2016oai,2015Ap&SS.359...57A,Chanda:2019hyh,Debnath:2019yge,Ilijic:2018ulf}.

We can conclude that the  comparison of our exact solution with the physical parameters of  pulsars gives indications that the model can realistically represent observed systems.  Furthermore, the approach can be extended to a large class of metrics and anisotropies, if the above physical requirements are satisfied. However, a detailed confrontation with observational data is needed.  This will be the argument of a forthcoming paper.

\section*{Acknowledgments}

 SC acknowledges the support of  INFN ({\it iniziative specifiche}  MOONLIGHT2 and QGSKY).
This paper is based upon work from COST action CA15117 (CANTATA), supported by COST (European Cooperation in Science and Technology). The authors want to thank the anonymous referee for the useful suggestions that  allowed to improve the paper.

\bibliography{JRPHSRefSalva}

\end{document}